\documentclass{amsart}
\usepackage{amsmath,amsthm,amssymb}
\usepackage{subfig}
\usepackage{graphicx,color}
\usepackage[margin=1.5in]{geometry}
\usepackage{amsrefs}
\usepackage{placeins}
\usepackage{adjustbox}

\renewcommand{\Im}{\mathrm{Im}~}
\newcommand{\Tr}{\mathrm{Tr}}

\newcommand{\mc}[1]{\mathcal{#1}}

\newcommand{\ud}{\,\mathrm{d}}

\newcommand{\averageop}[3]{\langle#1\lvert#2\rvert#3\rangle}

\newcommand{\braket}[2]{\langle#1\vert#2\rangle}

\newcommand{\Or}{\mathcal{O}}

\newcommand{\atom}{\text{atom}}

\newtheorem*{theorem*}{Theorem}

\newcommand{\pexsisigma}{PEXSI-$\Sigma$ }
\newcommand{\pexsizero}{vacuum boundary condition}

\newcommand{\REV}[1]{#1}

\title[PEXSI-$\Sigma$: a Green's function embedding method]{PEXSI-$\Sigma$: A Green's function embedding
method for Kohn-Sham density functional theory}

\author{Xiantao Li}
\address{Department of Mathematics, The
  Pennsylvania State University, University Park, PA 16802}
\email{xxl12@psu.edu}

\author{Lin Lin}
\address{Department of Mathematics, University of California, Berkeley
and Computational Research Division, Lawrence Berkeley National
Laboratory, Berkeley CA 94720 USA}
\email{linlin@math.berkeley.edu}

\author{Jianfeng Lu} 
\address{Department of Mathematics,
    Department of Physics, and Department of Chemistry, Duke
    University, Durham, NC 27708 USA}
\email{jianfeng@math.duke.edu}

\thanks{The work of X.L.~was supported by the National Science
  Foundation under award DMS-1522617. The work of L.L.~was partially
  supported by Laboratory Directed Research and Development (LDRD)
  funding from Berkeley Lab, provided by the Director, Office of
  Science, of the U.S. Department of Energy under Contract No.
  DE-AC02-05CH11231, the Alfred P. Sloan foundation, the DOE
  Scientific Discovery through the Advanced Computing (SciDAC) program
  and the DOE Center for Applied Mathematics for Energy Research
  Applications (CAMERA) program. The work of J.L.~was supported in
  part by the National Science Foundation under awards DMS-1312659 and
  DMS-1454939.}

\begin{document}

\begin{abstract}
% As Kohn-Sham density functional theory (KSDFT) being applied
%     to increasingly more complex materials, the periodic boundary
%     condition associated with supercell approaches also becomes
%     unsuitable for a number of important scenarios.  Green's function
%     embedding methods allow a more versatile treatment of complex
%     boundary conditions, and hence provide an attractive alternative
%     to describe complex systems that cannot be easily treated in
%     supercell approaches.  

\REV{In this paper, we propose a new
    Green's function embedding method called PEXSI-$\Sigma$ for
    describing complex systems within the Kohn-Sham density functional
    theory (KSDFT) framework, after revisiting the physics literature of
    Green's function embedding methods from a numerical linear algebra
    perspective.
    % We then propose a new Green's function embedding method called
    % PEXSI-$\Sigma$.
} The PEXSI-$\Sigma$ method approximates the density
  matrix using a set of nearly optimally chosen Green's functions
  evaluated at complex frequencies. For each Green's function, the
  complex boundary conditions are described by a self energy matrix
  $\Sigma$ constructed from a physical reference Green's function,
  which can be computed relatively easily. In the linear regime, such
  treatment of the boundary condition can be numerically exact.  The
  support of the $\Sigma$ matrix is restricted to degrees of freedom
  near the boundary of computational domain, and can be interpreted as
  a frequency dependent surface potential. This makes it possible to
  perform KSDFT calculations with $\mathcal{O}(N^2)$ computational
  complexity, where $N$ is the number of atoms within the
  computational domain. Green's function embedding methods are also
  naturally compatible with atomistic Green's function methods for
  relaxing the atomic configuration outside the computational domain.
  As a proof of concept, we demonstrate the accuracy of the
  PEXSI-$\Sigma$ method for graphene with divacancy and dislocation
  dipole type of defects using the DFTB+ software package.
% for efficient treatment of the boundary condition.  The
%$\Sigma$ matrices can be constructed using Green's functions
%corresponding to any reference system that shares the same potential in
%the exterior degrees of freedom. $\Sigma$  is only a surface potential
%and does not introduce additional interaction in the interior degrees of
%freedom.  Hence for systems with large interior degrees of freedom, the
%calculation can be performed efficiently using the recently developed
%pole expansion and selected inversion method (PEXSI). Numerical results
%using non-self-consistent DFTB+ calculations for water dimer, graphene
%with divacancy and graphene with dislocation dipole demonstrated the
%accuracy of the method. 
%
\end{abstract}

\maketitle

\section{Introduction}

%Outline
%Green's function. Multiscale analysis. Supercell method. Dense matrix.
%PEXSI. Extension as a embedding method. Extension with QM/MM. Proof of principle in DFTB+.
%Multiple defects. QM/MM materials.

\REV{This paper concerns the simulation of defects in materials in the
  framework of Kohn-Sham density functional theory
  (KSDFT)~\cite{HohenbergKohn1964,KohnSham1965}. Here we use the term
  ``defect'' to refer to general local perturbations such as
  vacancies, dislocations, in the otherwise smoothly deformed lattice
  structure in materials.  We are interested in cases that the global
  system is too large to be modeled entirely by KSDFT, so that we can
  only afford to ``embed'' the defect in an auxiliary system, in which
  the number of degrees of freedom is comparable to that of the defect region
  itself. In physics literature this procedure is known as
  ``embedding''. In the context of KSDFT, the goal of embedding is to
  correctly evaluate the density matrix corresponding to the defect
  region.  The simplest embedding scheme only includes the defect
  together with some nearby degrees of freedom, and places the resulting
  auxiliary system in vacuum. This scheme often leads to large error
  for real materials simulation.  Practically used embedding schemes
  often modify the degrees of freedom near the boundary of the
  auxiliary system to mimic the materials environment.  Analogous to
  the setup in partial differential equations (PDEs), we view such
  modification as a ``boundary condition''.  One common procedure is
  to embed the defects in a ``supercell'', so that the auxiliary
  system is periodically extended.  In the past two decades, the
  supercell approaches, such as those based on planewave basis
  sets~\cite{PayneTeterAllenEtAl1992,KresseFurthmuller1996,MakovPayne1995},
  have been the most widely used methods in computational material
  science to model defects. On the other hand, many systems are not
  periodic to start with, and the inherent periodic boundary treatment
  in supercell approaches is therefore not always suitable.  } Quantum
transport, defect migration, defect-defect interaction, and
dislocations are just a few examples of scenarios where the periodic
boundary condition encounters significant difficulties.
%and more versatile methods for describing complex boundary
%conditions are desirable.

%In order to model perturbations in materials (henceforth generally
%referred to as ``defects'') without enforcing periodic boundary
%conditions, it is desirable to ``embed'' the defects into the
%underlying environment, and to incorporate environmental effects by
%means of appropriate and more versatile boundary conditions. 
Various embedding
schemes~\cite{Cortona1991,KniziaChan2013,GoodpasterAnanthManbyEtAl2010,HuangPavoneCarter2011,
GarciaLuE:07,ChenOrtner2015}
have been developed in the literature in order to model complex
material structures more efficiently \REV{without using the periodic boundary
conditions}. Such methods allow the defect region to be
treated not only at the level of KSDFT, but also at higher levels of
\REV{electronic structure methods} such as the coupled cluster theory, though the
accuracy of the latter approach of embedding is significantly more
difficult to analyze from a numerical analysis perspective.
%such as those based on the
%density~\cite{Miller}, density matrix~\cite{Chan}, potential
%function~\cite{HuangCarter} and Green's
%functions
%have been   Among such schemes, 
In this paper we focus on Green's function
methods~\cite{BernholcLipariPantelides1978,ZellerDederichs1979,WilliamsFeibelmanLang1982,KellyCar92,
ZgidChan2011,KananenkaGullZgid2015,NguyenKananenkaZgid2016,ChibaniRenSchefflerEtAl2016,LinLuYingEtAl2009,LinChenYangEtAl2013},
and treat the defect region at the level of KSDFT.
\REV{Green's function methods evaluate the density matrix through the
linear combination of a number of Green's functions evaluated at complex
frequencies.} \REV{Since they} 
allow a more versatile treatment of
complex boundary conditions, \REV{they offer} an attractive
alternative to describe complex systems.  \REV{In particular, they have been successfully
applied to real materials simulation when localized basis functions are
available.}  
\REV{Examples of Green's function methods include the locally self-consistent
multiple scattering (LSMS)
method~\cite{WangStocksSheltonEtAl1995,NicholsonStocksWangEtAl1994},
Fermi operator expansion method~\cite{GoedeckerTeter1995, Goedecker:99}, 
the recent extension of the Korringa-Kohn-Rostoker (KKR)
method~\cite{ZellerDeutzDederichs1982,ZellerDederichsUjfalussyEtAl1995}
called KKRnano~\cite{ThiessZellerBoltenEtAl2012}, and the PEXSI
method~\cite{LinLuYingEtAl2009,LinChenYangEtAl2013}.  }

%However, a notable
%disadvantage of conventional Green's function methods for planewave
%basis sets is that they require the inversion of many dense matrices,
%which arises from discretization points of a contour in the complex
%plane. 
%Thus for large systems, localized basis sets are
%necessary. To overcome the computational cost of inverting large
%matrices, % The matrix inversion procedure
% becomes expensive for solving systems either with large number of
% atoms or discretized with large basis sets.
%Green's function methods have been combined with localization based
%methods to achieve linear scaling complexity for solving KSDFT for
%insulating systems. 

\vspace{1em}
\noindent\REV{\textbf{Contribution:}}

%The main contribution of our work is to propose a new Green's function
%embedding method called PEXSI-$\Sigma$, which is a natural extension of
%PEXSI to impose the boundary condition for the Green's functions in the
%defect region.  We consider local defects embedded in a physical
%reference system such as a crystal. Our assumption is that we can obtain
%the Green's function $G^0$ for the reference system (\textit{e.g.}, by
%means of a band structure calculation using the periodicity of the
%reference problem). The key idea of our approach is to obtain the
%boundary condition of the perturbed system via a Schur complement, which
%can be readily calculated by the reference Green's function $G^0$. If
%the problem is linear (\textit{e.g.}, for a model Hamiltonian without
%self-consistent field iteration), our scheme in fact provides the
%numerically exact density matrix  restricted to the defect region, and
%there is no error if the physical observable, such as the atomic force,
%only depends the local electronic structure. The modified boundary
%condition is a $\Sigma$ matrix that is only nonzero at the degrees of
%freedom near the surface of the defect region, and hence it can be
%interpreted as a surface potential, which might be embedding 

\REV{In this work, we consider defects embedded in a physical reference
system such as a crystal, modeled at the level of KSDFT. We assume the
Hamiltonian operator is discretized using a local basis set, and that we
can obtain a number of Green's functions $G^0(z)$ for the reference
system evaluated at different complex frequencies $z$. These reference
Green's functions can be obtained, for instance, by means of a band
structure calculation using the periodicity of the reference problem.
Then we propose a method to model the defects by an auxiliary system,
which contains the defect and a \textit{minimal} set of degrees of
freedom defined according to the sparsity of the Hamiltonian operator. 
The Hamiltonian operator for this auxiliary system is constructed by the
Hamiltonian operator of the global system restricted to the auxiliary
system, plus a frequency-dependent term that only modifies a submatrix
corresponding to boundary degrees of freedom. 
This extra term is closely related to a Schur complement, and can be
interpreted as a  discrete version of the Dirichlet-to-Neumann (DtN) map
operator for the global system~\cite{EngquistMajda1977,GivoliPatlashenkoKeller1998,KellerGivoli1989}. 
In physics literature, such modification
is a special type of ``self energy'' contribution. Following
standard notation in physics, we denote this extra term by
$\Sigma(z)$, where $z$ is a complex frequency at which the Green's
function $G(z)$ needs to be evaluated. We demonstrate that in the linear
regime, i.e., in the absence of self-consistent-field (SCF) iteration, our scheme
provides a numerically exact density matrix restricted to the defect
region. In such case, there is no error in computing physical observables such
as the atomic force in the defect region. 

Since $\Sigma(z)$ is only nonzero at the boundary of the auxiliary
system, we can efficiently evaluate $G(z)$ for the auxiliary system
using the pole expansion and selected inversion (PEXSI)
method~\cites{LinLuYingE2009,LinLuYingEtAl2009,LinYangMezaEtAl2011,LinChenYangEtAl2013}.
The computational complexity of the PEXSI method is at most
$\Or(N^{2})$, where $N$ is the number of atoms within the computational
domain.  The PEXSI method does not rely on the near-sightedness
principle~\cite{Kohn1996}, but only relies on the sparsity of the
Hamiltonian matrix. Hence the PEXSI method is applicable to metallic systems at
room temperature. The PEXSI method can be scalable on massively parallel
computers~\cite{JacquelinLinYang2015,JacquelinLinWichmannEtAl2015}.
PEXSI has been integrated into a number of electronic structure software
packages such as
SIESTA~\cite{LinGarciaHuhsEtAl2014,SolerArtachoGaleEtAl2002},
BigDFT~\cite{MohrRatcliffBoulangerEtAl2014},
CP2K~\cite{VandeVondeleKrackMohamedEtAl2005} and
DGDFT~\cite{LinLuYingE2012,HuLinYang2015a}, and has been used for
accelerating materials simulation with $10000$ atoms or
more~\cite{HuLinYangEtAl2014,HuLinYangEtAl2016}. 

PEXSI is a Green's function method for solving KSDFT for the global
system, and our development can be naturally combined with the PEXSI
method, which is referred to as the PEXSI-$\Sigma$ method.  The $\Sigma(z)$
modification introduced in this work only modifies matrix elements of
the Hamiltonian corresponding to boundary degrees of freedom, which
allows us to solve the auxiliary system with at most
$\Or(N^{2})$ cost, where $N$ is the number of atoms in the auxiliary
system.  We also present how to combine the PEXSI-$\Sigma$ method
seamlessly with atomistic Green's function methods~\cite{Li2009b,Li2012}
for structural relaxation of the defect system. 
 
As a proof of concept, we implement the PEXSI-$\Sigma$ method in the
DFTB+ software package~\cite{AradiHourahineFrauenheim2007}, and
demonstrate the accuracy using a water dimer, graphene with divacancy,
and graphene with a dislocation dipole with relaxed geometric structure
without SCF iterations. Our numerical results indicate that the
PEXSI-$\Sigma$ method can obtain accurate description
of the energy and forces in the defect region.}

\vspace{1em}
\noindent\REV{\textbf{Related work:}}

\REV{In physics literature, the ``self energy'' matrix (or $\Sigma$ matrix)
has been used in the context of} the non-equilibrium
Green's function (NEGF) method in quantum transport
calculations (e.g. ~\cite{BrandbygeMozosOrdejonEtAl2002}). 
\REV{Both the PEXSI-$\Sigma$ approach and the NEGF approach modify the
boundary degrees of freedom through Schur complements, but there are
important differences. In the context of modeling local defects in a
crystal, the strategy in the NEGF approach would require the Green's function
corresponding to a crystal but with the defect region removed.
The resulting system resembles a crystal with a ``hole'' corresponding
to the defect region, and this unphysical system can be very difficult to solve. On
the other hand, PEXSI-$\Sigma$ only requires the knowledge of Green's
functions for the physical crystal configuration, and such Green's
functions are much easier to compute.
In fact, we think our strategy for
constructing $\Sigma$ matrices could be potentially beneficial in the context of
quantum transport calculations as well for certain systems.
Another type of Green's function embedding methods use the Dyson
equation (e.g.~\cite{WilliamsFeibelmanLang1982,KellyCar92}), which uses 
physical reference Green's functions. However, the Dyson equation
requires dense linear algebra to be performed over the entire auxiliary
system, and the computational cost is therefore $\Or(N^3)$, where $N$ is
the number of atoms in the auxiliary system.
Meanwhile, PEXSI-$\Sigma$ only modifies the Hamiltonian matrix
corresponding to boundary degrees of freedom and is hence more
efficient.  } 
Our method 
is also related to the embedding method proposed by
Inglesfield~\cite{Inglesfield1981}, which is based on matching the boundary condition
for each individual eigenfunction. This strategy could be viable when
eigenfunctions are well separated from each other \REV{spectrally}. However, when
eigenfunctions are clustered such as for large scale KSDFT calculations,
it becomes impractical to derive the boundary condition for each
%supercell. 
eigenfunction.
%Mathematically, our approach is closely related to the
%Dirichlet-to-Neumann (DtN) map, which has been
%successfully obtained or approximated in various
%contexts~\cite{EngquistMajda1977,GivoliPatlashenkoKeller1998,KellerGivoli1989}.
%To the extent of our
%knowledge, the construction of such DtN type operator has not been
%presented in the electronic structure literature.  Furthermore, we
%extend the Green's function embedding method to structural relaxation
%using the recently developed atomic Dirichlet-to-Neumann
%scheme~\cite{Li2012}.
%\LL{Mention [ChenOrtner] somewhere?}

%These approaches do not directly address the multiscale coupling
%problem. We plan to develop approaches that are generally applicable to
%a large range of discretization schemes. Transport. 

%Many strategies have been developed in the literature to overcome this
%problem.\LL{May need a number of Green's function embedding techniques
%and fragment based embedding decomposition schemes; Also KKR type of methods..}

We note that the spirit of Green's function embedding methods are very different
from that of the quantum mechanics / molecular mechanics (QM/MM)
method, which is widely used in chemistry and
biology~\cite{WarshelLevitt1976}.  In the QM/MM method,
the coupling of the two types of models is usually a significant challenge. 
While QM models involve the degrees of freedom associated with
electrons (for example, electron density or electron orbital
functions), MM models do not explicitly take into account those
degrees of freedom.  One intuitive way to understand the issue at the
boundary is that the decomposition of the domain into QM and MM
regions creates ``dangling bonds'' at the interface. Therefore, a
popular approach is to introduce hydrogen-type atoms to passivate
those bonds.  More advanced approaches have been proposed to further
reduce the artifacts introduced by the coupling. See for example the
review articles~\cite{GaoTruhlar:02, LinTruhlar:07, SennThiel:09,
BrunkRothlisberger:15}.  We remark that the bond passivation model becomes 
challenging in materials science simulations, such as the description
of a local defect in aluminum. 
In Green's function embedding methods, the
coupling is through the boundary conditions imposed on the Green's
function of the QM domain, rather than changing the local chemical
environment of the coupling region. In particular, no bond passivation
is required.

\vspace{1em}
\noindent\REV{\textbf{Organization:}}

The manuscript is organized as follows. We briefly introduce Kohn-Sham
density functional theory and Green's function methods in
section~\ref{sec:green}, and review existing
Green's function methods from a numerical linear algebra perspective
in section~\ref{sec:embed}. \REV{In section~\ref{sec:newmethod} we introduce
a new Green's function method called PEXSI-$\Sigma$, and a geometry
relaxation method based on atomistic Green's functions.} We report the numerical results
using DFTB+ in section~\ref{sec:numer}, and discuss future directions in
section~\ref{sec:conclusion}.

%\section{Embedding: Green's function perspective}\label{sec:green}
\section{Preliminaries}\label{sec:green}

In Kohn-Sham density functional theory, the ground-state electron charge
density ${\rho}(x)$ of an atomistic system can be obtained from the
self-consistent solution to the Kohn-Sham equations
\begin{equation}
  \widehat{H}\left[{\rho}\right] \psi_i(x) = \psi_i(x) \varepsilon_i,
\label{eqn:kseqs}
\end{equation}
where $\widehat{H}$ is the Kohn-Sham Hamiltonian that depends on $\rho$, and 
$\{\psi_i(x)\}$ are the Kohn-Sham orbitals. The Kohn-Sham orbitals in turn determine the
charge density by
\begin{equation}
  {\rho}(x) = \sum_{i=1}^{\infty} |\psi_i(x)|^2 f_i.
\label{rhodef}
\end{equation}
The occupation numbers $\{f_i\}$ are chosen according to the Fermi-Dirac
distribution function
\begin{equation}
f_i= f_{\beta} (\varepsilon_i - \mu) = \frac{2}{1+e^{\beta(\varepsilon_i-\mu)}},
\label{fermidirac}
\end{equation}
where $\mu$ is the chemical potential chosen to ensure that
\begin{equation}
  \int {\rho}(x) \ud x = N_e.
\label{chargesum1}
\end{equation}
$\beta$ is the inverse  temperature, i.e.,
$\beta = 1/(k_B T)$ with $k_B$ being the Boltzmann constant and $T$ the temperature.
\REV{The nonlinear iteration with respect to the electron density
$\rho$ can be carried out using a self-consistent-field iteration
(SCF) procedure~\cite{RMartin}.}

The electronic-structure problem can be recast in terms of the
one-particle density matrix defined by 
\begin{equation}
\widehat{\Gamma}=\sum_{i=1}^{\infty}
|\psi_{i}\rangle f_{\beta}(\varepsilon_i-\mu) 
\langle \psi_i| = f_{\beta} (\widehat{H} - \mu), 
\label{eqn:gammaeq}
\end{equation}
and the chemical potential $\mu$ chosen so that
$\Tr\, \widehat{\Gamma} = N_e$, which is exactly the same constraint as \eqref{chargesum1}.

To solve for $\rho$ or $\widehat{\Gamma}$ in practice, we may choose a
finite basis set $\{\varphi_j\}$, and use a
Galerkin approximation for \eqref{eqn:kseqs} as the generalized
eigenvalue problem
\begin{equation}
H[\rho] C = S C \Lambda,
  \label{eqn:geneig}
\end{equation}
where $H_{ij}=\averageop{\varphi_{i}}{\widehat{H}}{\varphi_{j}}$ is the
projected Hamiltonian matrix, and
$S_{ij}=\braket{\varphi_{i}}{\varphi_{j}}$ is the overlap matrix. 
The matrix representation of the density matrix, denoted by $\Gamma$,
can be obtained from the generalized eigenvalue
decomposition~\eqref{eqn:geneig} as
\begin{equation}
    \Gamma = C f_{\beta}(\Lambda-\mu) C^{T}.
    \label{eqn:matrixgamma}
\end{equation}
For simplicity we consider the case when real arithmetic is used, and $H,
S, \Gamma$ are real symmetric matrices. The extension to the complex
Hermitian case is straightforward. Using linear algebra notation, let us denote by $\Phi=[\varphi_1,\cdots,\varphi_{N}]$
the matrix collecting all $N$ basis functions. Then 
the density matrix in the real space can be compactly approximated by
%is correspondingly approximated by  
%\LL{Some ambiguity here since $f$ is not a function of $H-\mu S$}
\begin{equation}
    \widehat{\Gamma} \approx \Phi \Gamma \Phi^{T}.
%    f_{\beta}(H - \mu S) = \Phi \Gamma \Phi^{T},
\end{equation}
%where $\Phi=[\varphi_1,\cdots,\varphi_{N}]$ is the matrix collecting all
%$N$ basis functions, and $\Gamma$ is the matrix representation of the
%density matrix. 

%Standard Green's function using the contour \dots PEXSI changes it to
%\dots \LL{Maybe a comparison of the contour that is used in standard
%Green's function embedding methods for PEXSI.}

It turns out that, in KSDFT calculations with the local density
approximation (LDA) and generalized gradient approximation (GGA)
for the exchange-correlation functionals, not all entries of the one-particle
density matrix are needed. In order to carry out the self-consistent
field iteration, it is sufficient to compute the electron density $\rho$, the diagonal
entries of $\widehat{\Gamma}$ in the real space, i.e.,
\begin{equation}
    {\rho}(x) \approx \Phi(x) \Gamma \Phi^T(x)
     = \sum_{ij}\Gamma_{ij}\varphi_{j}(x)\varphi_{i}(x).
  \label{eqn:rho}
\end{equation}
When the basis functions $\varphi_{i}(x)$ are compactly supported in
real space, the product of two functions $\varphi_i(x)$ and
$\varphi_j(x)$ \REV{is zero} when they do not overlap. This leads to
sparse Hamiltonian matrix $H$ and overlap matrix $S$, respectively. It also
implies that in order to compute $\rho(x)$, we only need $\Gamma_{ij}$
such that $\varphi_{j}(x)\varphi_{i}(x)\ne 0$ in
Eq.~\eqref{eqn:rho}. As shall be seen later, such sparsity plays a key
role in our method. 

The Kohn-Sham equations~\eqref{eqn:kseqs} are well-defined for closed
systems such as systems in vacuum (i.e., with Dirichlet boundary
condition imposed far away from the system) and with periodic boundary
condition. However, the eigenvalue formulation imposes major
difficulty for treating {\em open} systems. For instance, the
embedding of a defect into a crystalline system, which can be a point
defect such as a vacancy, or a line defect such as a dislocation. As
opposed to the solution of PDEs where tailored boundary conditions can
be formulated for specific operators such as in the case of the
absorbing boundary condition~\cite{EngquistMajda1977}, in KSDFT each
eigenfunction satisfies a different PDE, and hence requires its own
tailored boundary condition. The number of eigenfunctions is
proportional to the number of electrons $N_{e}$.  Finding such
boundary conditions is not only expensive when $N_{e}$ becomes large,
but also may not be a stable procedure since the eigenvalues of
interest are often clustered, or even form continuous energy bands in
the thermodynamic limit for solid state systems.

Here we demonstrate that the one-particle density matrix can serve as
a useful tool for quantum embedding. First, $\widehat{\Gamma}$ can be
evaluated without the need for diagonalization, if the Fermi function
is approximated by a linear combination of a number of simpler
functions.  This is the idea behind the Fermi operator expansion (FOE)
method~\cite{Goedecker1993}.  The FOE method is typically used as a
linear scaling method to accelerate KSDFT calculations for insulating
systems with substantial band gaps, or for general systems under very high
temperature. The recently developed pole expansion and selected
inversion (PEXSI) method extends the FOE method by means of an
efficient rational approximation, and significantly accelerates KSDFT
calculations for large scale metallic systems at room
temperature~\cites{LinLuYingE2009,LinLuYingEtAl2009,LinYangMezaEtAl2011,LinChenYangEtAl2013}.
% The PEXSI method does not rely on the near-sightedness
% principle~\cite{Kohn1996}, but only relies on the sparsity of the
% $H,S$ matrices. 

% The PEXSI method is particularly effective for
% low-dimensional quantum systems, and can solve Kohn-Sham equations for
% more than $10000$ atoms, and it has been integrated into a number of
% electronic structure software packages such as
% SIESTA~\cite{LinGarciaHuhsEtAl2014,SolerArtachoGaleEtAl2002},
% BigDFT~\cite{MohrRatcliffBoulangerEtAl2014},
% CP2K~\cite{VandeVondeleKrackMohamedEtAl2005} and
% DGDFT~\cite{LinLuYingE2012,HuLinYang2015a}.
%
\begin{figure}[h]
  \begin{center}
      \includegraphics[width=0.40\textwidth]{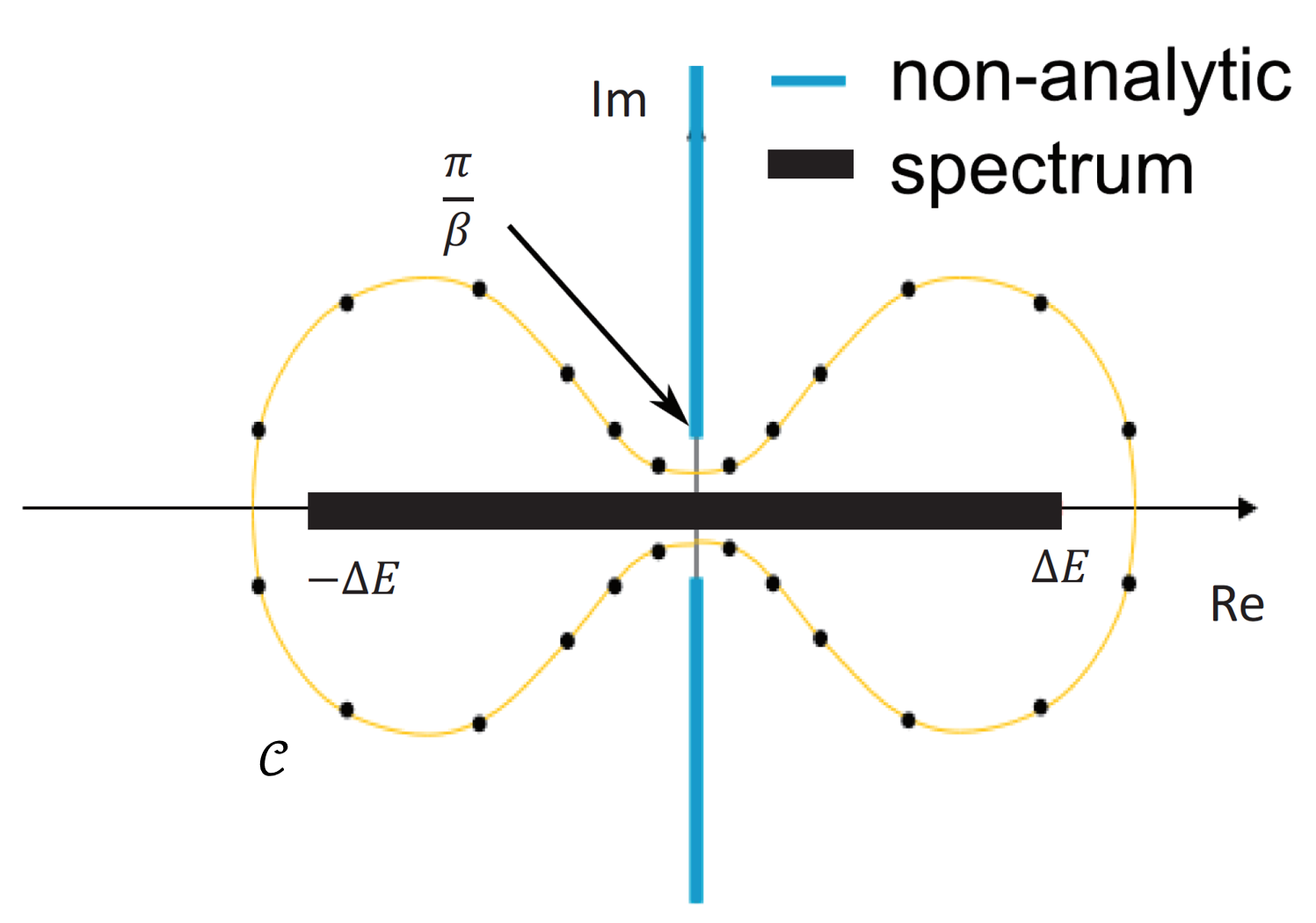}
  \end{center}
  \caption{Sketch of the contour used in the PEXSI method.}
  \label{fig:contour}
\end{figure}

In the PEXSI method, the single particle density matrix can be exactly
reformulated by means of a contour integral as
\begin{equation}
    \widehat{\Gamma}(x,x') = \frac{1}{2\pi i} \oint_{\mathcal{C}} f_{\beta} (z -
    \mu) (z - \widehat{H})^{-1}(x,x') \ud z.
    \label{eqn:contour}
\end{equation}
Here $\mathcal{C}$ 
%{\color{blue} this has been used for the density matrix, may be $\omega$?}
can be any contour that encircles the spectrum of $\widehat{H}$ without
enclosing any pole of the Fermi-Dirac function.
In the pole expansion~\cite{LinLuYingE2009}, we carefully choose a contour as in
Fig.~\ref{fig:contour}, and approximate the
single particle density matrix $\widehat{\Gamma}$ by its $P$-term 
approximation, denoted by $\widehat{\Gamma}_{P}$ as
\begin{equation}
  \begin{split}
    \widehat{\Gamma}_{P}(x,x') &= \Phi(x) \Im\left(
    \sum_{l=1}^{P}\frac{\omega^{\rho}_l}{(z_l+\mu) S - H}\right)
    \Phi^T(x')\\
    &\equiv \Phi(x) \Gamma_{P} \Phi^T(x').
  \end{split}
  \label{eqn:gammapole}
\end{equation}
The complex shifts $\{z_{l}\}$ and weights $\{\omega^{\rho}_l\}$ are
determined only by $\beta,\Delta E$ (the spectrum width of the matrix
pencil $(H,S)$) and the
number of poles $P$. These coefficients are known explicitly and their
calculation takes negligible amount of time. The pole expansion is an
effective way for approximating the one-particle density matrix, since
it requires only $\Or(\log \beta\Delta E)$ terms of simple rational
functions. 
%Note that the contour integral is not
%unique. In particular, any contour encircling the spectrum with proper
%treatment of the residues (if any) can be used. 
With some
abuse of notation, in the following discussion we will drop the
subscript $P$ originating from the $P$-term pole expansion
approximation unless otherwise noted. 

Eq.~\eqref{eqn:gammapole} converts the problem of computing
the one-particle density matrix by means of eigenfunctions into a
problem of evaluating $P$ inverse matrices or {\em Green's functions}, defined as
\begin{equation}
  G_{l} = \bigl((z_{l}+\mu)S-H\bigr)^{-1}, \quad l=1,\ldots,P.
  \label{eqn:green}
\end{equation}
Note that in order to evaluate the electron density, we only need to
evaluate the entries $(G_{l})_{ij}$ such that $H_{ij},S_{ij}\ne 0$.
\REV{This allows the PEXSI method to compute such selected elements of
an inverse matrix efficiently. We will discuss more along this line in
section~\ref{sec:pexsisigma}.} 

%It would seem that the need to carry out $P$ matrix inversions in
%(\ref{eqn:gammapole}) would mean that the computational complexity of
%this approach is still close to the $\Or(N^3)$ scaling of
%diagonalization. 
%However, the recently developed 
%provides an efficient way of computing the selected elements of an
%inverse matrix. For a (complex) symmetric matrix of the form $A=H-zS$,
%the selected inversion algorithm first constructs an $LDL^T$
%factorization of $A$, where $L$ is a block lower triangular matrix
%called the Cholesky factor, and $D$ is a block diagonal matrix. The
%computational scaling of the selected inversion algorithm is only
%proportional to the number of nonzero elements in the Cholesky factor
%$L$,  which is $\Or(N)$ for quasi-1D systems, $\Or(N^{1.5})$ for 
%quasi-2D systems,
%and $\Or(N^{2})$ for 3D bulk systems, thus achieving universal
%asymptotic improvement over the diagonalization method for systems of all
%dimensions.  It should be noted that the selected inversion algorithm is
%an \textit{exact} method for computing selected elements of $A^{-1}$
%if exact arithmetic is employed, and in practice the only source of
%error originates from the roundoff error.

\section{\REV{Existing} Green's function embedding schemes}\label{sec:embed}

\REV{In the context of embedding, we only need to find the ``boundary
conditions'' for Green's functions $\{G_l\}_{l=1}^{P}$. As mentioned in the
introduction, here the term ``boundary condition'' can refer to a
general way of modifying the degrees of freedom in an auxiliary system
to mimic the effects of the materials environment.} Since $P$ is
independent of the system size $N_{e}$, this
becomes a solvable problem even for systems of large sizes. On the other
hand, finding proper boundary conditions for $\Or(N_e)$ eigenvalue
problems can become impractical for systems of large
sizes~\cite{Inglesfield1981}. 
In this section we first review some existing ideas in the literature,
written in consistent linear algebra notation as used in the previous
section. 

\begin{figure}[h]
  \begin{center}
    \includegraphics[width=0.5\textwidth]{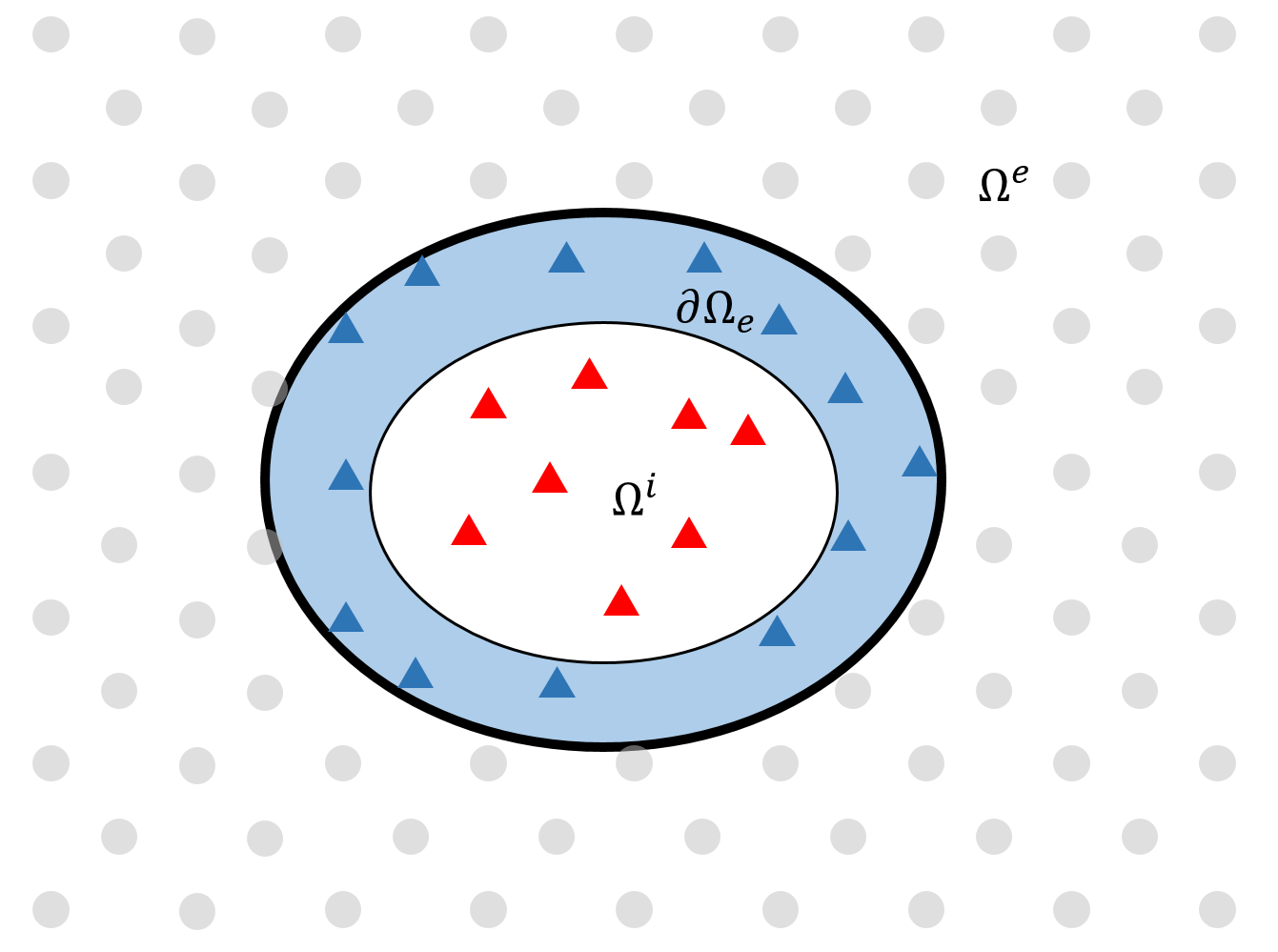}
  \end{center}
  \caption{Partition of the atoms in the computation domain into
  interior domain $\Omega^{i}$ (red triangles), boundary between
  interior and exterior domain $\partial \Omega^{e}$ (blue triangles in
  shaded area), and the rest of the exterior domain
  $\Omega^{e}\backslash \partial \Omega^{e}$
  (gray circles). 
  }
  \label{fig:partition}
\end{figure}

% Recall that our goal is to solve $\{G_l\}$ defined in \eqref{eqn:green}
% with finite dimensional $H$ and $S$ matrices.
In the embedding scheme,
we partition the degrees of freedom (i.e., nodal values associated with
the basis functions) into  
interior degrees of freedom $\Omega^{i}$ and exterior degrees of freedom $\Omega^{e}$, where
$\Omega^{i}\cap \Omega^{e}=\emptyset$. \REV{
In this paper we assume atom-centered basis functions are used in
discretizing the Hamiltonian operator. This \REV{type of basis set
includes} atomic orbitals, Gaussian type orbitals, as well as the
density-functional tight binding (DFTB) approximation~\cite{AradiHourahineFrauenheim2007}, which will be used in
our numerical examples. With some abuse of notation, we aggregate
degrees of freedom corresponding to the single atom, as illustrated in
Fig.~\ref{fig:partition}, and perform the partition geometrically
according to atomic positions. 
% When discrete basis functions $\Phi$ such as atomic
% orbitals are used, each basis function is associated with an atom. 
%Hence with some abuse of notation,
We will also not distinguish between the domain, and the set of
indices for the basis functions associated with the atoms in the
domain.  For example $H_{\Omega^{i},\Omega^{i}}$ represents the
diagonal matrix block of the Hamiltonian matrix for the basis
functions associated with atoms in $\Omega^{i}$. 
Parts of $\Omega^{e}$ are labeled as boundary degrees of freedom, denoted by
$\partial\Omega^{e}$, which is defined to be the collection of indices
$k$ so that $H_{\Omega^{i},k}\ne 0$.
As a result $H_{\Omega^{i},\Omega^{e} \backslash \partial\Omega^{e}}=0$.
Hence $\partial\Omega^{e}$ 
defines the \textit{minimal separation} between the defect and the environment in the
algebraic sense.} We partition $H$ accordingly into the block form
\REV{\begin{equation}
  \begin{split}
  H =& 
  \left(
  \begin{array}{l l|l}
    H_{\Omega^{i},\Omega^{i}} & H_{\Omega^{i},\partial\Omega^{e}} &
    0\\
    H_{\partial\Omega^{e},\Omega^{i}} & H_{\partial\Omega^{e},\partial\Omega^{e}} &
    H_{\partial\Omega^{e},\Omega^{e}\backslash \partial\Omega^{e}}\\
    \hline
    0 & H_{\Omega^{e}\backslash \partial\Omega^{e},\partial\Omega^{e}} &
    H_{\Omega^{e}\backslash \partial\Omega^{e},\Omega^{e}\backslash \partial\Omega^{e}}
  \end{array}
  \right)\\
  \equiv &
  \left(
  \begin{array}{l l|l}
    H_{\alpha\alpha} & H_{\alpha\beta} &
    0\\
    H_{\beta\alpha} & H_{\beta\beta} &
    H_{\beta 2}\\
    \hline
    0 & H_{2\beta} &
    H_{22}
  \end{array}
  \right)
  \equiv 
  \left(
  \begin{array}{c|c}
      \marginbox{0.85em}{$H_{11}$}  & H_{12} \\ 
    \hline
    H_{21}  & H_{22}
  \end{array}
  \right).
  \label{eqn:Hpartition}
  \end{split}
\end{equation}
For convenience of the discussion in the sequel, we introduce the
short hand notation
$\alpha\equiv \Omega^{i},\beta\equiv \partial \Omega^{e}$, and
$1\equiv \Omega^{i}\cup \partial \Omega^{e} \equiv \alpha \cup \beta$
and $2\equiv\Omega^{e}\backslash \partial \Omega^{e}$.}
%\REV{For convenience of the discussion later, we have defined as well the
% index $1\equiv \Omega^{i}\cup \partial \Omega^{e}$ and
%$2\equiv\Omega^{e}\backslash \partial \Omega^{e}$, respectively. }
%\jl{please check} 
Other matrices of the same size, such as the overlap matrix
$S$ and the density matrix $\Gamma$, can be partitioned accordingly. As
will be seen below, grouping $\Omega^i$ and $\partial \Omega^e$ together
allows accurate calculation of local physical quantities such as atomic
forces corresponding to the degrees of freedom in $\Omega^i$. 

The \REV{atomic configuration} in $\Omega^{i}$ can be fully
disordered and/or \REV{involve} defects, but we assume that the atomic
configuration in $\Omega^{e}$ is not far away from relatively simple
configurations, such as crystalline systems for which the Green's
function can be evaluated or approximated \REV{using a band structure
calculation, which is not expensive compared to the cost of evaluating
the global system with defects.}
The quantity of interest is the density matrix restricted to
$\Omega^{i}$. To this end we need to evaluate $\Gamma_{11}$.  We also
require an embedding scheme to result in \REV{a discretized system in
  the basis $\Phi$ involving only degrees of freedom in
  $1\equiv \Omega^{i} \cup \partial \Omega^e$, and the information from the
  rest of the domain will be incorporated implicitly.} 

Below we omit the subscript $l$ (the index of the poles), and denote by 
\[
A = (z+\mu)S - H, \quad \text{and} \quad G=A^{-1}.
\]
Note that the $z$ dependence is implicit in the
notation. The submatrices of $G$ satisfy the equation
\begin{equation}
  \begin{pmatrix}
    A_{11} & A_{12}\\
    A_{21} & A_{22}
  \end{pmatrix}
  \begin{pmatrix}
    G_{11} & G_{12} \\
    G_{21} & G_{22} 
  \end{pmatrix} 
  = 
  \begin{pmatrix}
    I_{1} & 0  \\
    0 & I_{2} 
  \end{pmatrix}, 
  \label{eqn:greenid}
\end{equation}
where $I_{1},I_{2}$ are identity matrices. 

Green's function embedding methods typically involve two atomic
configurations. We denote by $H^{0}$ and $S^{0}$ the matrices
corresponding to a reference system, and $H$ and $S$ the matrices
corresponding to a physical system of interest. For simplicity we
assume that after discretization, the dimension of $H^{0}$ and $H$ are
the same.  This assumption is clearly violated when atoms are added or
removed from the systems. However, this condition can be relaxed in
the practical numerical schemes as illustrated in
section~\ref{sec:pexsisigma}.  We also assume that the reference
\REV{density matrix} and the physical \REV{density matrix} can be
evaluated using the same contour \REV{using Eq.~\eqref{eqn:gammapole}}
, and define
\[
A^{0} = (z+\mu)S^{0} - H^{0}.
\]
In physical terms, this means that we choose the same chemical potential
for the two systems. In this paper we assume the reference atomic
configuration is a perfect crystal. In the presence of localized defect,
it is possible to use such grand canonical ensemble treatment with fixed
chemical potential. However, for finite sized reference systems, the
grand canonical treatment is only an approximation, and \REV{updating} the
chemical potential to adjust for the correct number of electrons may become necessary.

\subsection{Schur complement method}

The most straightforward way to reduce the degrees of freedom in
$\Omega^{e}$ is via the use of a Schur complement (a.k.a Gaussian
elimination). The Schur complement method eliminates the $A_{22}$
submatrix directly, and obtain 
\begin{equation}
  \left(A_{11} + \Sigma\right) G_{11}
  = 
  I_{1}.
  \label{eqn:greenschur}
\end{equation}
%\begin{equation}
%  \begin{pmatrix}
%    A_{11} & A_{12} \\
%    A_{21} & S
%  \end{pmatrix}
%  \begin{pmatrix}
%    G_{11} & G_{12} \\
%    G_{21} & G_{22}
%  \end{pmatrix}
%  = 
%  \begin{pmatrix}
%    I_{1} & 0 \\
%    0 & I_{2}
%  \end{pmatrix}.
%  \label{eqn:greenschur}
%\end{equation}
Here
\begin{equation}
\Sigma =  - A_{12}A_{22}^{-1}A_{21}
  \label{eqn:Schur}
\end{equation}
is called the Schur complement, which reflects the impact of the
exterior degrees of freedom to the interior degrees of freedom.
\REV{We note that the use of $\Sigma$ to denote the Schur complement is
different from the convention in numerical linear algebra. We choose
this notation here and below due to the direct connection of Schur complement
and the ``self energy'' matrix in physics literature, which is often
denoted by $\Sigma$.}
The Schur complement $\Sigma$ depends on the complex shift $z$. 
In physics literature, $\Sigma$ is often
referred to as the self energy
matrix~\cite{Mahan2000,BrandbygeMozosOrdejonEtAl2002}.
%Compared to Eq.~\eqref{eqn:greenid}, \eqref{eqn:greenschur} involves
%the low rank update matrix as the second term in the parenthesis on the
%left hand side of the equation. 
%\jl{not sure if I understand the previous sentence}\LL{just removed it}
The matrix inverse $A_{22}^{-1}$ can be interpreted as the Green's
function corresponding to a physical system with only degrees of freedom
in $\Omega^{e}\backslash \partial \Omega^{e}$. In Fig.~\ref{fig:partition} this corresponds to the
degrees of freedom represented by
gray circles, which is a system containing a very
large void by excluding the degrees of freedom in
$\Omega^{i}\cup \partial \Omega^{e}$. In
term of the reference system, the corresponding reference matrix $A^{0}$
takes the form
\[
A^{0} = 
  \begin{pmatrix}
    0 & 0 \\
    0 & A_{22}
  \end{pmatrix}.
\]

For quasi-one-dimensional systems, the Schur complement method
has been successfully applied in first principle quantum transport
calculations using the non-equilibrium Green's function
methods~\cite{BrandbygeMozosOrdejonEtAl2002}.
In such calculations, the vacancy system becomes two
independent semi-infinite systems, and can be calculated efficiently by
means of recursive Green's function
methods~\cite{Lopez-SanchoLopez-SanchoRubio1984}. This technique
becomes very costly for systems in two and three dimensions, since the cost
of computing $A_{22}^{-1}$ can be similar to that of the computation of
the entire system. 
%Furthermore, the system
%involving only exterior degrees of freedoms may easily contain dangling
%bonds, and its solution may encounter difficulty in obtaining
%self-consistent solutions of the nonlinear Kohn-Sham equations. \jl{I am not sure if I understand this comment ... one problem we should mention is perhaps that introducing such a vacancy might shift the chemical potential a lot; so that the contour of the two systems will be different, if we insist on that point of view}

\subsection{Dyson equation method}

To overcome the above mentioned difficulty associated with the Schur complement
method, let us consider more general reference systems, with the requirement that they only differ with $A$ in the $A_{11}$ block, i.e., 
\begin{equation}
  \Delta A \equiv A^{0} - A = 
  \begin{pmatrix}
    A^{0}_{11}-A_{11} &  0\\
    0  & 0 
  \end{pmatrix}.
  \label{eqn:Adiff}
\end{equation}
Nonetheless, %denote by $G^{0}=(A^{0})^{-1}$, 
even local changes in $A_{11}$ can lead to extended changes in terms of
the difference of Green's functions $G-G^{0}$.  Green's function
embedding methods can be regarded as approximations to solutions of
$G_{11}$ without the explicit involvement of the rest of blocks. 
%\jl{I
%removed $G_{12}$ here which seems to be remainder of some previous
%version of the algorithm}

One possible way to achieve this is described by Williams, Feibelman and
Lang~\cite{WilliamsFeibelmanLang1982}, and later extended by Kelly and
Car~\cite{KellyCar92}, through the Dyson's equation. Again using the same
numerical linear algebra notation, here we demonstrate that the
Dyson equation method can be interpreted equivalently using the
Sherman-Morrison-Woodbury formula.  \REV{The Dyson's equation 
can be derived by starting with
\( (A^0 - \Delta A) G =I,\) and left multiplying the equation by $G^0$, which yields,
\[ (I-G^0 \Delta A ) G = G^0.\]
This is typically rewritten as,}
\begin{equation}
  G = G^{0} + G^{0} \Delta A G, 
  \label{eqn:dyson}
\end{equation}
or equivalently
\[
G = (I-G^{0}\Delta A)^{-1} G^{0}.
\]
We view $\Delta A$ as a ``low-rank update'' and rewrite as
\[
\Delta A = E_{1}(\Delta A)_{11}E_{1}^{T},
\]
where $E_{1}^{T}=[I_{1}, 0]$. Then by the Sherman-Morrison-Woodbury formula,  we have,
\[
G = G^{0} + G^{0} E_{1} (\Delta A)_{11} (I_{1} -
G^{0}_{11} (\Delta A)_{11} )^{-1}
E_{1}^{T} G^{0}.
\]
In order to evaluate the electron density in $\Omega^{i}$, it is sufficient to evaluate
$G_{11}$ as
\begin{equation}
  G_{11} = G^{0}_{11} + G^{0}_{11} (\Delta A)_{11} (I_{1} -
  G^{0}_{11} (\Delta A)_{11} )^{-1}
  G^{0}_{11}.
  \label{eqn:kellycar}
\end{equation}
Note that all quantities, including the matrix inverse in
Eq.~\eqref{eqn:kellycar} only involves matrices restricted to the
degrees of freedom in $\Omega^i \cup \partial \Omega^e$, and \REV{the results from
Eq.~\eqref{eqn:kellycar} and ~\eqref{eqn:greenschur} are equivalent}.

Compared to the Schur complement approach, one advantage of the
Dyson equation approach is that the reference system can be chosen to be physically more
meaningful for systems of all dimensions.  In particular, for configurations such as the one in 
Fig.~\ref{fig:partition}, Green's functions corresponding to the
crystalline configuration can be efficiently
computed by means of a band structure calculation, and 
can be readily used in Eq.~\eqref{eqn:kellycar}.
%This
%procedure is also numerically more stable due to the absence of dangling
%bonds.

Another advantage of the Dyson equation approach is that physical quantities,  such
as the differences of energy between the physical system of interest and
the reference system can be evaluated accurately, even for the
contribution to the energy differences in $\Omega^{e}$.
To see why this is possible, we first note that in the contour integral
formulation, physical quantities, such as the total number of
electrons and total energy can be computed
with the trace of differences of Green's functions, multiplied by the
overlap matrix, i.e., $\Tr[SG-S^{0}G^{0}]$. 
%\jl{What is our assumption on $S$ and $S^0$? Should it be $SG - S^0 G^0$?} Using Eq.~\eqref{eqn:kellycar}, 
%\begin{equation}
%  \Tr[S (G-G^{0})] = \Tr\left[(I_{1} -
%  (G^{0})_{11} (\Delta A)_{11} )^{-1}
%  (G^{0}SG^{0})_{11} (\Delta A)_{11}\right].
%  \label{}
%\end{equation}
%Here the difficult term is $(G^{0}SG^{0})_{11}$, which
%involves $G^{0}_{12}$ and is not available. However, note that
Note that both $G$ and $G^{0}$ are $z$-dependent, and we have the identity
\[
\Tr[G S] = \frac{d}{dz} \Tr\left[ \log (zS-H) \right] = \frac{d}{dz}\log \det (zS-H),
\]
and similarly
\[
\Tr[G^{0}S^{0}] = \frac{d}{dz} \log \det (zS^{0}-H^{0}).
\]
Here we used the identity $\Tr[\log (\cdot)] = \log[\det(\cdot)]$.
%Then
%\begin{multline}
%    \frac{d}{dz} \Tr\left[ \log\left( I_{1} -
%  (G^{0})_{11} (\Delta A)_{11} \right) \right]\\
%  = \Tr\left[(I_{1} -
%  (G^{0})_{11} (\Delta A)_{11} )^{-1}
%  (G^{0}SG^{0})_{11} (\Delta A)_{11}\right].
%  \label{eqn:tracelog}
%\end{multline}
%We have
%\[
%\Tr[S (G-G^{0})] = \frac{d}{dz} \Tr\left[ \log\left( I_{1} -
%  (G^{0})_{11} (\Delta A)_{11} \right) \right] 
%  = \frac{d}{dz} \log\left[ \det\left( I_{1} -
%  (G^{0})_{11} (\Delta A)_{11} \right) \right].
%\]
Then we have %\jl{note that I changed $I$ to $I_1$ in the last term, in consistency with other places}
\begin{equation}
\begin{aligned}
  \Tr[G S] - \Tr[G^{0}S^{0}] =  &\frac{d}{dz} \log \det
  (G^{0}G^{-1}) \notag \\
  =&\frac{d}{dz}\log\det(I-G^{0}\Delta A) = 
  \frac{d}{dz}\log\det(I_1-G^{0}_{11}(\Delta A)_{11}),
  \label{eqn:tracelog}
\end{aligned}
\end{equation}
where we have used Dyson's equation for $G^{0}G^{-1}$.
In order to compute differences of energy, free energy or number of
electrons, only the determinant of matrices restricted to
$\Omega^{i} \cup \partial \Omega^{e}$ is needed.  In
practice  the $\frac{d}{dz}$ operator can be approximated using a finite difference scheme in the complex
plane. 

\REV{Although the reference Green's function $G^{0}$ can be 
efficiently computed by means of a band structure calculation,
} the disadvantage of the Dyson equation approach is that the matrix
$G^0_{11}$ in Eq.~\eqref{eqn:kellycar} is a dense matrix. Hence dense
linear algebra must be used for matrix-matrix multiplication and matrix
inversion operations. The computational cost can still be large when a
large \REV{number of} degrees of freedom in $\Omega^{i}$ is needed.

\section{\REV{A new Green's function method}}\label{sec:newmethod}

\subsection{The PEXSI-$\Sigma$ method}\label{sec:pexsisigma}

Let us
now %The setup in section~\ref{sec:previous} allows us to readily
introduce the PEXSI-$\Sigma$ method, which is our new strategy of
treating the boundary conditions for the Green's function. 

We first note that $A^{0}$ and $A$ only differ in the $A_{11}$ block
as in Eq.~\eqref{eqn:Adiff}, and 
\REV{the Schur complement in Eq.~\eqref{eqn:Schur} can be either given by
the reference system or the defect system, i.e.}
\begin{equation}
  \Sigma =  - A_{12}A_{22}^{-1}A_{21} =
- A_{12}^{0}(A^{0}_{22})^{-1}A^{0}_{21}.
  \label{}
\end{equation}
\REV{Consequently, $\Sigma$ as in Eq.~\eqref{eqn:greenschur} can also be
defined using $G^{0}$ as}
\[
G^{0}_{11} (A^{0}_{11}+\Sigma) = I_{1},
\]
or equivalently
\begin{equation}
  G^{0}_{11}\Sigma = I_{1} - G^{0}_{11} A_{11}^{0}. 
  \label{eqn:dtn1}
\end{equation}
Here we demonstrate that Eq.~\eqref{eqn:dtn1} can be used to give a
compact representation for $\Sigma$. 
\REV{Recall that in Eq.~\eqref{eqn:Hpartition} we split the collective index $1$ into
$(\alpha,\beta)\equiv (\Omega^{i},\partial \Omega^{e})$. Then
Eq.~\eqref{eqn:Schur} can be written as
}
\begin{equation}
  \Sigma = -A_{12}A_{22}^{-1}A_{21} = 
  \REV{
  \begin{pmatrix}
    0 \\ A_{\beta 2}
  \end{pmatrix}}
  A_{22}^{-1}\REV{
  \begin{pmatrix}
    0 &
    A_{2 \beta} 
  \end{pmatrix} }
  \equiv
  \begin{pmatrix}
    0 & 0\\
    0 & \Sigma_{\beta\beta}
  \end{pmatrix}.
  \label{}
\end{equation}
\REV{Therefore the $\Sigma$ matrix is only nonzero on the diagonal
matrix block corresponding to $\beta\equiv \partial\Omega_{e}$.} 
Then Eq.~\eqref{eqn:dtn1} can be
written as
\begin{equation}
  \begin{pmatrix}
    G^{0}_{\alpha\alpha} & G^{0}_{\alpha\beta} \\
    G^{0}_{\beta\alpha} & G^{0}_{\beta\beta}
  \end{pmatrix}
  \begin{pmatrix}
    0 & 0 \\
    0 & \Sigma_{\beta\beta}
  \end{pmatrix}
  = 
  I_{1} - G^{0}_{11} A_{11}^{0}.
  \label{eqn:dtn2}
\end{equation}
Here we have used the fact that $\Sigma$ only has non-zero component on the
boundary degrees of freedom. Take the $(\beta,\beta)$ component
of the equation~\eqref{eqn:dtn2}, and we have
\begin{equation}
G^{0}_{\beta\beta} \Sigma_{\beta\beta} =
I_{\beta}-G^{0}_{\beta\alpha}A^{0}_{\alpha\beta} -
G^{0}_{\beta\beta}A^{0}_{\beta\beta},
  \label{eqn:dtn0}
\end{equation}
or in a more compact form
\begin{equation}
  \Sigma_{\beta\beta} =  (G_{\beta\beta}^{0})^{-1}(I -
  G^{0}_{\beta\alpha}A_{\alpha\beta}^{0}) - A_{\beta\beta}^{0}.
  \label{eqn:dtn}
\end{equation}

\REV{Compared to previous schemes in section~\ref{sec:embed}}, our
approach has the following advantages: 1) It is an accurate
reformulation of the embedding scheme under the same assumption of the
non-zero pattern of $\Delta A$ as that in the Dyson equation approach.
Hence the reference Green's function $G^{0}$ can correspond to a
physical reference system, such as the crystalline configuration.  2)
\REV{Compared to the Dyson equation approach, the advantage of using
  Eq.~\eqref{eqn:dtn} is that it introduces a modification matrix {\it
    only} on the boundary degrees of freedom $\partial \Omega^{e}$,
  and hence the reduced system remains to be a sparse system for
  systems of large sizes.}  \REV{This is crucial for using fast
  methods such as PEXSI, of which the effectiveness relies on the
  sparsity of the matrix $A$.}

%Such 
%sparsity structure ensures that the computational complexity is still at
%most $\Or(N^2)$ where $N$ is the number of degrees of freedom corresponding to
%$\Omega^{i}\cup \partial \Omega^{e}$.

\REV{ More specifically, for a symmetric matrix of the form $A=zS-H$,
  the selected inversion
  algorithm~\cite{LinLuYingEtAl2009,LinYangMezaEtAl2011,JacquelinLinYang2015}
  first constructs an $LDL^T$ factorization of $A$, where $L$ is a
  block lower diagonal matrix called the Cholesky factor, and $D$ is a
  block diagonal matrix.  In the second step, the selected inversion
  algorithm computes all the elements $A^{-1}_{ij}$ such that
  $L_{ij}\ne 0$.  Since $L_{ij}\ne 0$ implies that
  $H_{ij},S_{ij}\ne 0$, all the required selected elements of $A^{-1}$
  are computed, and the computational scaling of the selected
  inversion algorithm is only proportional to the number of nonzero
  elements in the Cholesky factor $L$.~\cite{LinLuYingEtAl2009}.}  For
a finite size system, the size of this matrix block is approximately
the same as the number of degrees of freedom corresponding to the
surface of the system. \REV{Regarding the implementation, we can use
  the techniques in sparse linear algebra, and} reorder the matrix
$A=zS-H$ so that the interior degrees of freedom $\Omega^{i}$ appear
before the boundary degrees of freedom $\partial \Omega_{e}$. The
$\Sigma$ matrix only modifies the matrix block corresponding to
degrees freedom in $\partial \Omega_{e}$. This matrix block becomes
dense anyway, since it is the last block in the Gaussian elimination
procedure (or $LDL^{T}$ factorization)~\cite{LinLuYingEtAl2009}.  Therefore if number of
degrees of freedom in $\Omega^{i}$ is sufficiently large, the
modification due to $\Sigma$ only increases the prefactor of the
asymptotic complexity of selected inversion, which is at most
$\Or(N^2)$ and $N$ is the number of degrees of freedom corresponding
to $\Omega^{i}\cup \partial \Omega^{e}$.

With $G_{11}$ computed, physical observables that rely on the
local density matrix, such as the atomic force, can be readily
computed. In PEXSI, the Hellmann-Feynman force associated with the
$I$-th atom is given by~\cite{SolerArtachoGaleEtAl2002}
\begin{equation}
    F_{I}  = -\Tr\left[ \Gamma \frac{\partial H}{\partial R_I} \right]
    +\Tr\left[ \Gamma^E \frac{\partial S}{\partial R_I} \right].
  \label{eqn:forcepole}
\end{equation}
Analogous to the density matrix~\eqref{eqn:matrixgamma}, $\Gamma^{E}$ is
the energy density matrix defined by
\begin{equation}
  \Gamma^{E}=C \Lambda f_{\beta}(\Lambda-\mu)  C^T.
  \label{eqn:gammaE}
\end{equation}
\REV{It has been shown~\cite{LinChenYangEtAl2013}} that the energy density matrix can be computed using
the same set of Green's function $G_{l}$ as required for the density
matrix, but with different weights $\{\omega_{l}^{E}\}$
\begin{equation}
  \Gamma^{E} \approx \Im \left(C \sum_{l=1}^{P} \frac{\omega^{E}_l}{\Lambda-(z_l+\mu) I}
  C ^T\right) = \Im\left(\sum_{l=1}^{P}
  \frac{\omega^{E}_l}{(z_l+\mu)S-H}\right).
  \label{eqn:redEgamma}
\end{equation}
Note that the sparsity pattern of $\frac{\partial H}{\partial
R_I},\frac{\partial S}{\partial R_I}$ is the same as that of
$H,S$ respectively. Therefore if $I$ corresponds to an atom in
$\Omega^{i}$, the trace in Eq.~\eqref{eqn:forcepole} can be computed
using $\Gamma,\Gamma^{E}$ restricted to $\Omega^{i}\cup \partial
\Omega^{e}$, which is readily computed in the PEXSI-$\Sigma$
formulation.

In order to evaluate the energy or the number of electrons in the
global domain, one needs to either use exterior degrees of freedom
explicitly, or to use the approach described in
Eq.~\eqref{eqn:tracelog} for Dyson's equation, which we will not go into details here. On the other hand, PEXSI-$\Sigma$
can be immediately used to evaluate the number of electrons restricted
to $\Omega^{i}$, denoted by $N_{e}^{i}$, which is a useful quantity to
measure in charge transfer processes. Note that the global number of
electrons can be computed as $N_{e}=\Tr[S \Gamma]$, the interior
number of electrons can be computed as
\begin{equation}
    N_{e}^{i}= \Tr[S_{\alpha\alpha}\Gamma_{\alpha\alpha}] +
    \Tr[S_{\alpha\beta}\Gamma_{\beta\alpha}].
    \label{eqn:Nei}
\end{equation}
Similarly one can measure the interior band energy
\begin{equation}
    E_{\mathrm{band}}^{i}=
    \Tr[S_{\alpha\alpha}\Gamma^{E}_{\alpha\alpha}] +
    \Tr[S_{\alpha\beta}\Gamma^{E}_{\beta\alpha}],
    \label{eqn:Ebandi}
\end{equation}
which is the contribution of the total band energy
$E_{\mathrm{band}}=\Tr[S\Gamma^{E}]$ from the interior degrees of
freedom.

%Another useful physical observable that can be computed locally is the
%interior band energy
%\begin{equation}
%    
%    \label{}
%\end{equation}

%Compared to the Dyson equation method in~\eqref{eqn:tracelog},
%the PEXSI-$\Sigma$ method cannot evaluate quantities such as the
%global number of electrons and the global energy, without using exterior
%degrees of freedom explicitly.
%
%However, 
%
%discussed in section~\ref{sec:pexsisigma}, \jl{should we define this
%  here or in the previous section?} the ``interior band energy'',
%denoted by $E_{\mathrm{band}}^{i}$, should agree well between PEXSI
%and \pexsisigma.  

\subsection{Geometric relaxation by atomistic Green's function}
Another appealing aspect of the present approach is that the relaxation of the nuclei can be formulated within the same framework. 
In molecular mechanics, in order to predict structural properties of lattice defects, the surrounding atoms have to be relaxed so that
the system reaches a mechanical equilibrium. In principle, the forces on every atom can be computed
based on the Hellmann-Feynman theorem. With the same observation that
away from the defects, the lattice deformation is small, we linearize the atomic interaction in the exterior region. This standard approximation is known as the harmonic approximation \cite{AsMe76}, under which the force balance can be expressed
as  a linear system of finite difference equations, \REV{
\begin{equation}\label{eq:fd}
  f_I \equiv \sum_J D_{I,J} u_J=0, \qquad \forall\, I \in \Omega^e,
\end{equation}}
subject to boundary conditions from the interior region. 
Here $D_{I,J}$ is the force constant matrix corresponding to the
periodic lattice structure, defined as the second derivative of the
energy. In the context of QM/MM coupling, such approximation has also
been used in \cite{ChenOrtner2015}.  The force constant
matrix $D_{I, J}$ can be computed by means of a finite difference
approach (also called the ``frozen phonon approach''),
or by density functional perturbation
theory~\cite{BaroniGironcoliDalEtAl2001} in electron
structure software packages.  Since they are defined for a crystalline
structure, a supercell can be used for this purpose.  Similar to the
sparsity of the matrices $H$ and $S$, we will make a truncation for
$D_{I,J}$ based on the magnitude of the matrix, and denote the spatial
cutoff by $r_\text{cut}.$ \REV{An example will be given in the next section to illustrate how the truncation is done.}
% \jl{need to talk about truncation scheme here for the force constant
% matrix}
Notice that here we have assumed a same partition
of the domain into $\Omega^i$ and $\Omega^e$ as in the electronic
part. However, depending on the truncation radius, the sparsity of $D$
might be different 
compared to the Hamiltonian matrix $H$. Therefore we denote the boundary  by  $\partial \Omega^e_{\atom}$, as opposed to  the definition of  the boundary  for the electron part, which was denoted by $\partial
\Omega^e$.

Let us now  show that similar to the electron part, the atomic relaxation
can be determined using a more efficient  procedure so that atomic
degrees of freedom can be restricted to the boundary.
To see how this reduced model is derived, we use the matrix
representation and denote $ u_{\alpha}$, $u_{\beta}$, and $u_2$ the
displacement in the inner region $\Omega^i$, outer boundary $\partial
\Omega^e_{\atom}$ and exterior $\Omega^e \backslash\partial
\Omega^{e}_{\atom}$, respectively. 

%Then the force balance equations
%\eqref{eq:fd} can be rewritten as 
%\begin{equation}\label{eqn:Du}
%   D_{\beta\beta} u_\beta + D_{\beta2} u_2 =  -D_{\beta\alpha} u_\alpha,
%   \quad D_{2\beta} u_\beta + D_{22} u_2 =  0.
%\end{equation}
\REV{
Given $u_\alpha$, the atom displacement in the interior,  we are left to determine $u_\beta$ and $u_2$.
Our goal is to eliminate $u_2$, in order to remove the large number of degrees of freedom in the exterior domain. 
In analogy to Eq.~\eqref{eqn:Hpartition}, the force balance
equation~\eqref{eq:fd} can be rewritten as}
\begin{equation}
  \left(
  \begin{array}{l l|l}
    D_{\beta\alpha} & D_{\beta\beta} &
    D_{\beta 2}\\
    \hline
    0 & D_{2\beta} &
    D_{22}
  \end{array}
  \right)
  \begin{pmatrix}
    u_{\alpha}\\
    u_{\beta}\\
    u_{2}
  \end{pmatrix} = 
 \begin{pmatrix}
    0\\
    0 
  \end{pmatrix}.
  \label{eqn:blockatom}
\end{equation}
\REV{
From the partition of the domain, we have that 
$D_{\alpha2} = 0$.
Since Eq.~\eqref{eq:fd} is only valid for the indices,
$\Omega^{e}\equiv \beta\cup 2$, Eq.~\eqref{eqn:blockatom} has only two
row blocks.
Similar to $G^{0}$ for the electronic degrees of freedom, we define the
atomistic Green's function $\mc{G}=D^{-1}$. After eliminating the
degrees of freedom with respect to $u_{2}$ in Eq.~\eqref{eqn:blockatom},
we have}
\begin{equation}
  D_{\beta\alpha} u_{\alpha} +
  (D_{\beta\beta}+\Sigma^{\mc{G}}_{\beta\beta}) u_{\beta}=0.
  \label{eqn:betaatom}
\end{equation}
\REV{Here $\Sigma^{\mc{G}}_{\beta\beta}=-D_{\beta2} D_{22}^{-1}
D_{2\beta}$ is the Schur complement for the atomistic degrees of freedom.
Analogous to Eq.~\eqref{eqn:dtn0} we can obtain an equivalent formula
for $\Sigma^{\mc{G}}_{\beta\beta}$ using the physical reference Green's
function $\mc{G}$ as }
\begin{equation}
  \mc{G}_{\beta\beta} (D_{\beta\beta} + \Sigma^{\mc{G}}_{\beta\beta}) = I -
  \mc{G}_{\beta\alpha} D_{\alpha\beta}.
  \label{eqn:atomdtn0}
\end{equation}
\REV{Finally multiply $\mc{G}_{\beta\beta}$ to both sides of
Eq.~\eqref{eqn:betaatom} we have}
%\begin{equation}
%    \mc{G}_{\beta\alpha}
%    D_{\alpha\beta} + \mc{G}_{\beta\beta} D_{\beta\beta} + \mc{G}_{\beta2}
%    D_{2\beta}=I, \quad \mc{G}_{\beta\beta} D_{\beta2} + \mc{G}_{\beta2} D_{22}=0.
%    \label{}
%\end{equation}
%
%This simplifies Eq.~\eqref{eqn:Du} to
\begin{equation}\label{eq: bi0}
 u_\beta =  \mc{G}_{\beta\alpha}D_{\alpha\beta} u_\beta - \mc{G}_{\beta\beta} D_{\beta\alpha} u_\alpha.
\end{equation}
%\REV{This can be obtained by multiplying the first equation in \eqref{eqn:Du} by $\mc{G}_{\beta\beta} $ and the second equation
%in \eqref{eqn:Du} by $\mc{G}_{\beta2}$, and then solve for $u_\beta.$}

This forms a closed system for the displacement of the atoms at the boundary. The coefficients in this linear system
involve the force constant matrices and the Green's function for the reference state. Such equations have 
been derived and implemented in \cite{Li2009b,Li2012} as a
coarse-grained molecular mechanics model, \REV{and the derivation
presented in this work provides a unified perspective for Green's
function methods for electronic and atomic degrees of freedom.
Similar to the Green's function in the QM model, the atomistic Green's can be expressed as a Fourier integral in the first Brillouin zone. There are various techniques for
computing the Green's functions efficiently \cite{MaRo02,trinkle:014110}, especially when the interatomic distance is large.}

The geometric optimization can be obtained as follows: 
For the atoms in $\Omega^i$, the forces are determined from the KSDFT 
model, and the atomic positions are relaxed using a nonlinear solver, e.g., the conjugate-gradient
method.  These updated positions will be used as input in the Eq.~\eqref{eq: bi0}, which becomes a closed linear system for the displacement of the atoms in $\partial
\Omega^e_{\atom}$. Once the displacement along $\partial
\Omega^e_{\atom}$ is determined from \eqref{eq: bi0}, this equation
can be used to evaluate the displacement of the atoms that are further
out (e.g., those in $\Omega^e \backslash \partial
\Omega^e_{\atom}$).%\LL{Do you mean Eq. (23)?}.\jl{I think Eq. (31) is the correct reference}
%This second part is merely an evaluation step and costs even less.

Note that in this procedure the atomic degrees of freedom in
$\Omega^e$ are completely determined by those in $\Omega^{i}$. Due to our
choice of the reference system to be the periodic lattice for $A^0$,
in the current method, there is no feedback of the deformation of the
exterior domain to the $\Omega^i$. It would be an interesting future
direction to consider how to incorporate the change into the reference
Hamiltonian.

\section{Numerical results}\label{sec:numer}
  
In this section we demonstrate the accuracy of the \pexsisigma method
using three examples: a water dimer, a graphene system with a divacancy,
and a graphene system with a dislocation dipole with opposite Burgers
vectors under relaxed atomic configuration. Our method is implemented in
the DFTB+ code~\cite{AradiHourahineFrauenheim2007}. DFTB+ uses the density functional tight binding
(DFTB) method, which can be viewed as a numerical discretization of the
Kohn-Sham density equations with minimal degrees of freedom, and thus
allows the study of systems of relatively larger sizes without 
parallel implementation. DFTB+ defines a semi-empirical charge density,
which can be computed both self-consistently and non-self-consistently.
In the \pexsisigma method, self-consistent charge density calculation
requires the charge density in $\Omega^{e}$ to be
properly taken into account, which is not yet in the scope of this work.
Hence all calculations below are performed in the non-self-consistent
mode of DFTB+. In all calculations, the
electronic temperature is set to the room temperature $300$K. All
quantities are reported in atomic units (au) unless otherwise specified.
All the computation is performed on a single Intel i7
CPU processor with $64$ gigabytes (GB) of memory.

We report the results for the following methods. For the full system, we
compare the results from the exact diagonalization (DIAG) method and the pole expansion with 
selected inversion (PEXSI) method. We demonstrate that the results from
DIAG and PEXSI \REV{for the full system} fully agree with each other. 
\REV{We show the effectiveness of the \pexsisigma method without taking
into account directly the exterior degrees of
freedom.} \REV{As a proof of concept, the $\Sigma$ matrices are
constructed from PEXSI calculations for the reference system, and is then
fixed in the calculation with defects.}
In order to demonstrate the effectiveness
of the environment-dependent self energy matrix $\Sigma$, we also
compare with the results by setting $\Sigma$ to a zero matrix. This is referred to
as the \pexsizero method in this section. In the non-self-consistent
calculations, the \pexsizero method is equivalent to considering an
isolated system with the degrees of freedom in $\Omega^{e}$ directly
eliminated from the calculation. In all the examples, we find that the
inclusion of a properly approximated $\Sigma$ matrix significantly
improves the accuracy.

\subsection{Water dimer}\label{subsec:dimer}

\begin{figure}[h]
  \begin{center}
    \includegraphics[width=0.3\textwidth]{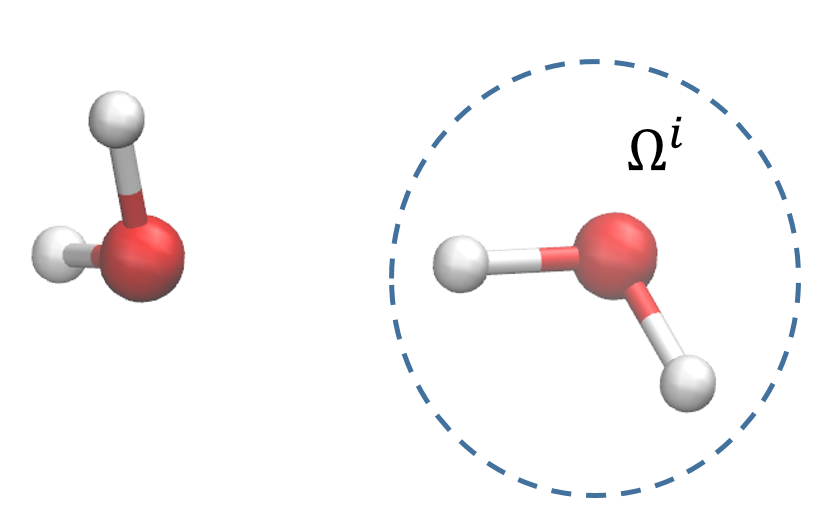}
  \end{center}
  \caption{Atomic configuration for water dimer. Large red ball: oxygen
  (O).
  Small white ball: hydrogen (H). The molecule in $\Omega^{i}$ is stretched along the
  O-O direction.
  }
  \label{fig:dimer}
\end{figure}

Our first example is a water dimer system (Fig.~\ref{fig:dimer}). The
system is partitioned into two parts, with one water molecule described
as $\Omega^{i}$ and the other molecule as $\Omega^{e}$.  Here $80$ poles are used in the PEXSI and \pexsisigma method to
guarantee accuracy. At the equilibrium configuration, the total energy
obtained from the DIAG method is $-8.1705870965$ au, and the total energy
obtained from the PEXSI method is $-8.1705870964$ au, with discrepancy
less than $10^{-10}$ au.  Therefore the results from DIAG and PEXSI
fully agree with each other.

In order to demonstrate that the \pexsisigma method gives accurate
results in different atomic configurations, we stretch the water
molecule in $\Omega^{i}$ along the oxygen-oxygen direction, and denote
by $\Delta d_{OO}$ the displacement away from equilibrium position.
In the \pexsisigma method, the \REV{value of the} Hamiltonian matrix elements between
$\Omega^{i}$ and $\Omega^{e}$ vary with respect to the change of the
atomic configuration. 
Hence in the absence of the energy contribution
from $\Omega^{e}$, the total energies obtained from PEXSI and
\pexsisigma in general do not agree with each other. 
However, as
discussed in section~\ref{sec:pexsisigma},
%\jl{should we define this here or in the previous section?} 
the interior band energy $E_{\mathrm{band}}^{i}$, together with the
atomic force corresponding to atoms in $\Omega^{i}$ should agree well between PEXSI
and \pexsisigma.  

Fig.~\ref{fig:dimer} (a), (b) report the interior
band energy, as well as the force on the oxygen atom in $\Omega^{i}$
projected along the O-O direction, respectively. We find that energies
and forces vary smoothly with respect to the change of the O-O
distance, and the results from PEXSI and \pexsisigma fully agree with
each other.  We remark that due to the small system size, the exterior
degrees of freedom $\Omega^{e}$ coincide with the boundary degrees of
freedom $\partial\Omega^{e}$.  Hence all $\Sigma$ matrices are
zero. In this special case, the \pexsisigma method and the \pexsizero
method are the same.

%In all calculations, the
%$\Sigma$ matrices in the \pexsisigma calculation are provided from Green's
%functions obtained from the crystalline configuration without defects.
%In all calculations, PEXSI uses $80$ poles to guarantee accuracy. The
%electronic temperature is set to the room temperature $300$ K.
%by studying two types of defects in graphene: divacancy
%and a dislocation dipole with opposite Burgers vectors.   

\begin{figure}[h]
  \begin{center}
    \includegraphics[width=0.45\textwidth]{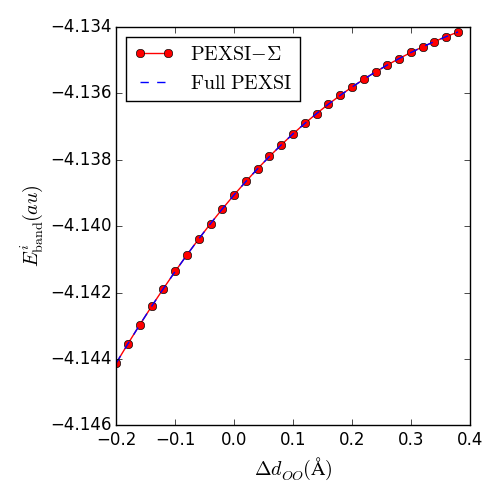}
    \quad
    \includegraphics[width=0.45\textwidth]{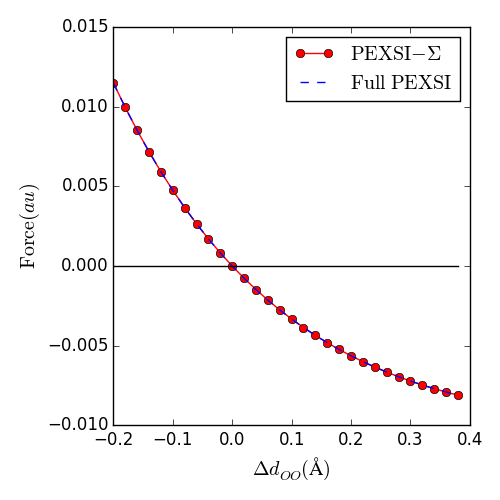}
  \end{center}
  \caption{(a) The interior band energy, and (b) the force on the
  oxygen atom in $\Omega^{i}$ projected along the O-O direction, as the
  molecule in $\Omega^{i}$ is stretched along the O-O direction,
  \REV{using PEXSI-$\Sigma$, and full simulation with PEXSI,
  respectively.  The horizontal line in (b) marks the equilibrium
  position ($\Delta d_{OO}=0$) for which the force vanishes.}
  }
\end{figure}

\subsection{Divacancy in graphene}\label{subsec:divacancy}

\begin{figure}[h]
  \begin{center}
    \includegraphics[width=0.4\textwidth]{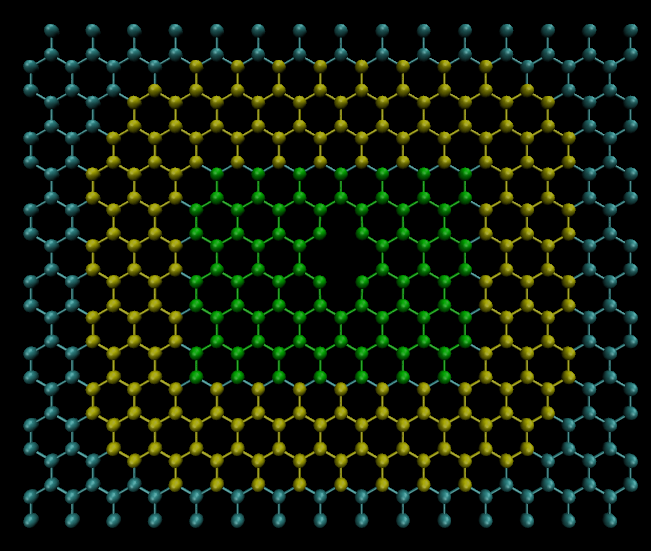}
  \end{center}
  \caption{Atomic configuration of the divacancy example in graphene
  with $418$ atoms, partitioned into interior
  atoms $\Omega^{i}$ (green),  boundary atoms $\partial\Omega^{e}$
  (yellow), and the rest of the exterior atoms $\Omega^{e}\backslash
  \partial\Omega^{e}$ (cyan). }
  \label{fig:divacancy1}
\end{figure}

Our second numerical example is a graphene system with a single
divacancy defect.  Starting from a periodic configuration with
$420$ atoms, two atoms are removed to create a divacancy
(Fig.~\ref{fig:divacancy1}). No further structural relaxation is
performed at this stage. In the periodic configuration without
the defect, the total energy computed from the DIAG method is
$-721.049897496$ au, and the total energy computed from the PEXSI
method with $80$ poles and at the same chemical potential is
$-721.049897489$ au. Hence the results from DIAG and PEXSI fully agree
with each other, and all numerical results below will be benchmarked
with that from the PEXSI method. 

\begin{figure}[h]
  \begin{center}
    \includegraphics[width=0.4\textwidth]{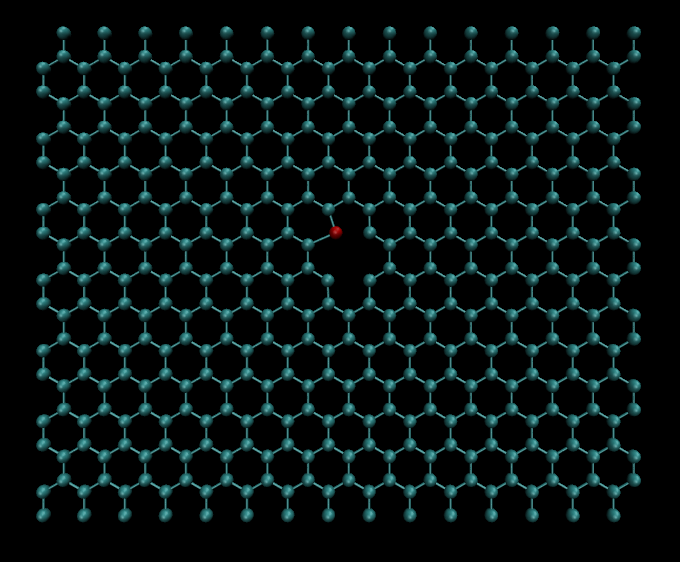}
  \end{center}
  \caption{Atomic configuration of the divacancy system with one atom (red)
  shifted by $0.1$ \AA{} along the $x$-direction.  The same partitioning
  strategy as in Fig.~\ref{fig:divacancy1} is used.}
  \label{fig:divacancymove}
\end{figure}

For the divacancy system, the atoms are partitioned according to
Fig.~\ref{fig:divacancy1}. Since $\Omega^{e}\backslash
\partial\Omega^{e}$ is non-empty, the $\Sigma$ matrices are non-zero. In
the \pexsisigma method, the $\Sigma$ matrices are obtained from the
PEXSI calculation in the periodic configuration. We compare the
interior band energy between the divacancy (D) and periodic
configuration (P) in Table~\ref{tab:ebandindivacancy}, obtained from
\REV{PEXSI for the full system, as well as from} \pexsisigma, and \pexsizero methods, respectively. In order to
assess the relative accuracy of the methods, we also compare the
interior band energy for another system by shifting one atom in
Fig.~\ref{fig:divacancy1} by a small distance of $0.1$ \AA\ along the
$x$-direction. The resulting configuration is denoted by SD
(shifted divacancy, Fig.~\ref{fig:divacancymove}). 

Table~\ref{tab:ebandindivacancy} indicates that in the periodic
configuration, the result from \pexsisigma fully agrees with that from
\REV{the simulation of the full system with PEXSI}.  Even though the
$\Sigma$ matrix is obtained from the periodic configuration, the
inclusion of $\Sigma$ matrices in the \pexsisigma formulation
significantly improves the accuracy in other atomic configurations as
well. The error of the energy
difference between the divacancy and periodic configuration using the
\pexsizero method is $0.0145$ au. This error is reduced by $67$ times to $0.0002$ au
in the \pexsisigma method. Similarly the error of the energy difference
between the divacancy and the shifted divacancy configuration using the
\pexsizero method is $0.0027$ au, and the error is reduced by about $75$ times
to $0.000036$ au in the \pexsisigma method.

%In order to use \pexsisigma method for geometry optimization, we also
We report the maximum of the error of the atomic forces calculated from
all interior atoms in Table~\ref{tab:errorforcedivacancy}. In
all configurations, the maximum force error obtained from the
\pexsisigma method is less than $3\times 10^{-5}$ au, which is very
accurate for geometry optimization and molecular dynamics studies. 
Compared to the \pexsizero method, the improvement due to the inclusion
of the $\Sigma$ matrix is again nearly 2 orders of magnitude.

\begin{table}
  \centering
  \begin{tabular}{l|c|c|c}
  \hline
   System & Full PEXSI & \pexsisigma & Vacuum\\
   \hline
  Periodic (P) & -145.70244 & -145.70244& -145.76624\\
  Divacancy (D) & -142.56345& -142.56367& -142.61273\\
  Shifted Divacancy (SD) & -142.45347& -142.45368&-142.50003\\
  \hline
  Energy difference (D-P) & 3.13899& 3.13877&3.15351\\
  Energy difference (SD-D) & 0.10999& 0.11003&0.11270\\
  \hline
  \end{tabular}
  \caption{The interior band energy for the graphene systems. Unit: au}
  \label{tab:ebandindivacancy}
\end{table}

\begin{table}
  \centering
  \begin{tabular}{l|c|c}
  \hline
   System & \pexsisigma & Vacuum\\
   \hline
  Periodic (P) &  0.00000 & 0.00407 \\
  Divacancy (D) & 0.00003 & 0.00399 \\
  Shifted Divacancy (SD) & 0.00003 & 0.00384 \\
  \hline
  \end{tabular}
  \caption{Maximum error of the force for interior degrees of freedom
  for the graphene systems. Unit: au}
  \label{tab:errorforcedivacancy}
\end{table}

\subsection{Dislocation dipole in graphene}\label{subsec:dipole}

%\begin{figure}[htp]
%  \begin{center}
%    \includegraphics[width=0.4\textwidth]{dislocation-figures/Dipole-atomview.png}
%  \end{center}
%  \caption{The atoms around the two dislocations.}
%  \label{fig:dipole}
%\end{figure}

\begin{figure}[htp]
  \centering 
  \includegraphics[width=0.3\textwidth]{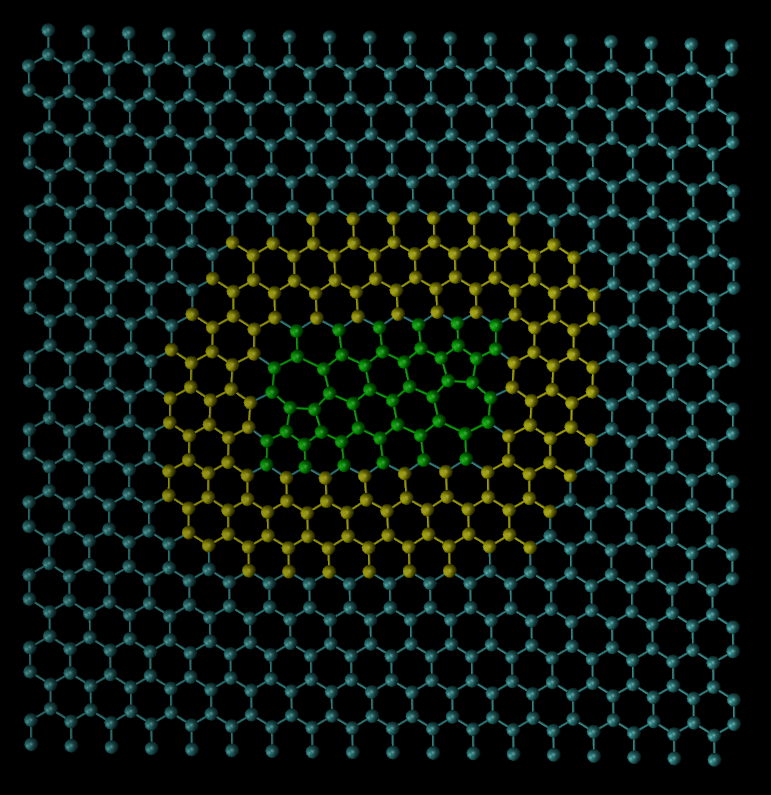}\qquad
   \includegraphics[width=0.41\textwidth]{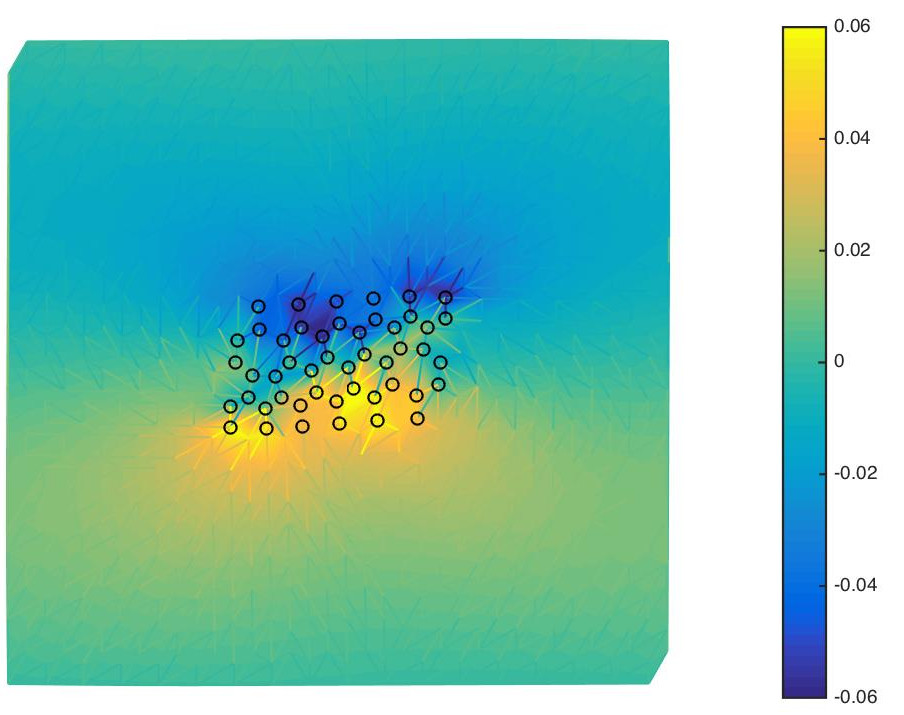}
  \caption{(Left) Atomic configuration of the dislocation dipole example 
  with $720$ atoms, partitioned into interior
  atoms $\Omega^{i}$ (green),  boundary atoms $\partial\Omega^{e}$
  (yellow), and the rest of the exterior atoms $\Omega^{e}\backslash
  \partial\Omega^{e}$
  (cyan). (Right) Displacement field (first component) after the
  geometric relaxation; the position of interior atoms is plotted on
  top. }
  \label{fig:dipole}
\end{figure}

\begin{figure}[htp]
  \begin{center}
    \includegraphics[width=0.4\textwidth]{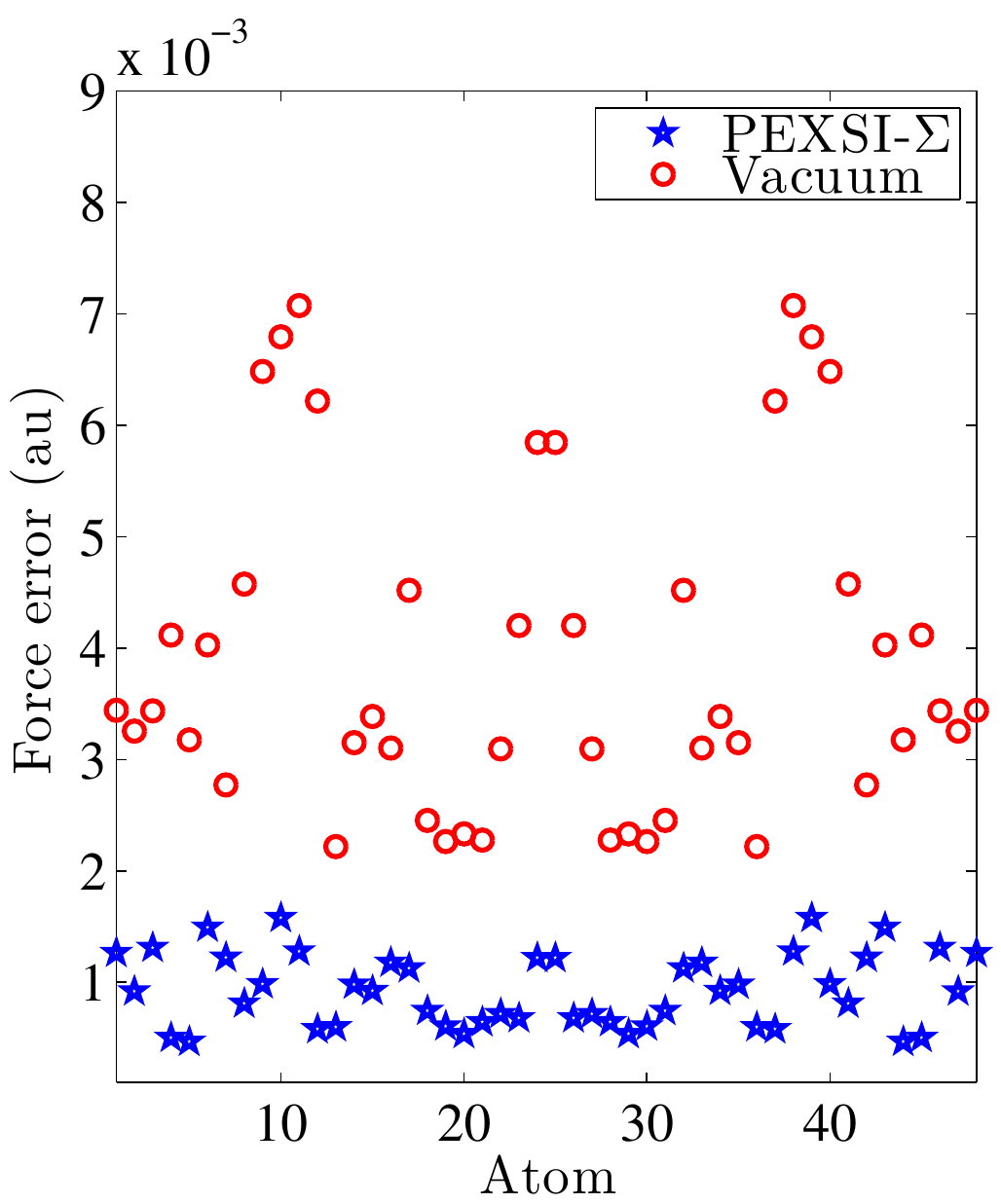}
  \end{center}
  \caption{Error of the atomic force for atoms in the interior domain of
  the dislocation dipole system.}
  \label{fig:dipole_force}
\end{figure}

In this test problem, we consider a dislocation dipole in the graphene
system. Such a dislocation can be identified as a
pentagon-heptagon (5-7) pairs among the hexagonal rings \cite{bonilla2012driving}. As
comparison, we form a supercell with 720 atoms in total. The entire
system is 4.55nm$\times$4.38nm. \REV{The lattice constant is set to $a_0=1.4247$\AA$.$ For the force
constant matrix $D$, we performed a calculation in DFTB+ using a supercell with 48 atoms. The matrix 
$D$ is then produced by DFTB as the Hessian matrix. Based on the magnitude of each $3\times3$ block, which corresponds to the interaction of an atom with its neighbors, we make a truncation. In particular, the diagonal block has norm ($l_2$ norm) about 
%$0.7715$(Hartree/$\AA^2$)
$0.2160$ au. We keep the force constants from up to 6th neighbors. The distance is about $2a_0,$  where the norm of the force constant matrix has been reduced to about 
%$0.006$(Hartree/$\AA^2$).
$0.0017$ au.  }
 Fig.~\ref{fig:dipole} (a) shows the
atomic configuration as well as the partition of the system. \REV{We observe that the cut-off of the atoms interactions is slightly smaller than that of the QM model.} Compared
to the example in section~\ref{subsec:divacancy}, the interior domain
is reduced to be just around the dislocation dipole.   Structural relaxation
is also performed for the entire system so that all atoms, including
the atoms in the exterior domain, deviate from the equilibrium
position, as shown in Fig.~\ref{fig:dipole} (b). The $\Sigma$ matrix is
still constructed from the graphene system with periodic structure.
Fig.~\ref{fig:dipole_force} shows that even with a small interior domain
and deformed atomic configuration in the exterior domain, the accuracy
of PEXSI-$\Sigma$ reduces the error of the force uniformly for all atoms
in the interior domain to be around $10^{-3}$ au.

%\begin{figure}[htp]
%  \begin{center}
%    \includegraphics[width=0.4\textwidth]{dislocation-figures/dspl_full.jpg}
%    \includegraphics[width=0.4\textwidth]{dislocation-figures/dspl_pexsi_sigma.jpg}
%  \end{center}
%  \caption{Comparison of the displacement: Left: results from full DFTB+ calculation; Right: results from PEXSI-SIGMA.}
%  \label{fig:disp}
%\end{figure}

%Figure. \ref{fig:disp} shows the atomic strain with $e_{11}$ and $e_{22}$ components. These local values are calculated using
%the Hardy's formulation, by writing a continuous representation $u(x)= \sum_i u_i \varphi(x-x_i),$ with $\varphi$ being a kernel function, here chosen as the spline kernel function with cut-off distance $R_{cut}= 5 \AA$. The strain then is defined from the gradient. 
% 
%\begin{figure}[htp]
%  \begin{center}
%    \includegraphics[width=0.42\textwidth]{dislocation-figures/strain_full.jpg}
%    \includegraphics[width=0.42\textwidth]{dislocation-figures/strain_pexsi_sigma.jpg}
%  \end{center}
%  \caption{Comparison of the displacement: Left: results from full DFTB+ calculation; Right: results from PEXSI-SIGMA.}
%  \label{fig:strain}
%\end{figure}

\section{Conclusion and future work}\label{sec:conclusion}
%Despite the large amount of work from 70s to 90s for the treatment of
%local defects by means of Green's function embedding methods, supercell
%type approaches become the dominant method today for solving electronic
%structure problems with local defects. However, Green's function methods
%remain to be an attractive alternative due to its capability for
%handling complex boundary conditions. In the context of local defects,
%Green's function embedding methods can in principle exactly embed the
%defect into an infinite crystal. On the other hand, current Green's
%function embedding methods often lead to dense matrix problems, and the
%computational cost can be large for systems with large interior degrees
%of freedom. 

In this work we proposed a new Green's function embedding method
called PEXSI-$\Sigma$ for efficient treatment of boundary
conditions in complex materials.  The $\Sigma$ matrices can be constructed using Green's
functions corresponding to any physical reference system that shares a
similar potential corresponding to exterior degrees of
freedom. The $\Sigma$ matrices can be viewed as a surface potential and
do not introduce
additional interaction among the interior degrees of freedom.  Hence
for systems with large number of interior degrees of freedom, the
calculation can be performed efficiently using the pole expansion and
selected inversion method (PEXSI). Numerical results using
non-self-consistent DFTB+ calculations for water dimer, graphene with
divacancy and graphene with dislocation dipole demonstrated the
accuracy of the method.

We note that our current implementation of the PEXSI-$\Sigma$ method,
which is only serial, is just a proof of
principle. % Our current implementation of
% PEXSI-$\Sigma$ is only only serial.
As indicated by the performance of the PEXSI
method~\cites{LinGarciaHuhsEtAl2014,JacquelinLinYang2015}, when the
number of interior degrees of freedom is large, the PEXSI-$\Sigma$
method should readily allow a massively parallel implementation in the
future with at most $\Or(N^2)$ complexity.  In order to apply the
PEXSI-$\Sigma$ method for the accurate computation of physical
quantities, we need to include the self-consistent field effect, which
requires the solution of a Coulomb-like equation on the global domain.
In particular, the electrostatic energy depends sensitively on the
total number of electrons in the system. It is most natural to use a
fixed chemical potential. This corresponds to the grand canonical
ensemble in the PEXSI-$\Sigma$ method, and may be a more natural
choice for describing processes with charge transfer. However, the
grand canonical ensemble treatment might need to be relaxed when the
reference system is of finite size.  The $\Sigma$ matrices
are constructed from $G^{0}$, which is only exact in the absence of
deformation of exterior degrees of freedom.  When the potential in the
exterior domain changes due to atomic relaxation or long range Coulomb
interaction, the correction to the $\Sigma$ matrix could be possibly
computed by means of perturbation theory. We also remark that Green's
function embedding methods may also become more versatile if the
$\Sigma$ matrices exhibit certain locality properties to accommodate
structural changes of atoms in the exterior domain such as in the
presence of a single dislocation, and also can be used to study
interaction of defects by using multiple disconnected QM regions.
Green's function embedding methods may also be an attractive
alternative for coupling with electronic structure theories beyond the
level of KSDFT (see e.g., the recent
works~\cite{ZgidChan2011,NguyenKananenkaZgid2016,ChibaniRenSchefflerEtAl2016}).
We plan to explore these directions in the future.

\bibliographystyle{amsxport}
\bibliography{dtnref}

\end{document}